\documentclass[journal]{IEEEtran}
\usepackage{amsthm}

\usepackage{fixltx2e}

\usepackage{graphicx}

\usepackage{subcaption}              

\usepackage[utf8]{inputenc}

\usepackage{amsfonts}
\usepackage{amsmath}
\usepackage{mathtools}

\usepackage{amssymb}

\usepackage{bm}

\usepackage{color,verbatim}
\usepackage{multirow}
\usepackage{accents}

\usepackage{theoremref}

\usepackage{url}

\newcounter{rulecounter}
\newcommand{\resetrule}{ \setcounter{rulecounter}{0}}
\resetrule

\newsavebox{\selvestebox}
\newenvironment{colbox}[1]
  {\newcommand\colboxcolor{#1}%
   \begin{lrbox}{\selvestebox}%
   \begin{minipage}{\dimexpr\columnwidth-2\fboxsep\relax}}
  {\end{minipage}\end{lrbox}%
   \begin{center}
   \colorbox{\colboxcolor}{\usebox{\selvestebox}}
   \end{center}}

\definecolor{orange}{rgb}{1,0.8,0}
\definecolor{gray}{rgb}{.9,0.9,0.9}
\definecolor{darkgray}{rgb}{.3,0.3,0.3}
\definecolor{darkblue}{rgb}{.1,0.0,0.3}
\definecolor{lightblue}{rgb}{0.7,0.7,1}
\definecolor{lightred}{rgb}{1,0.7,.7}
\definecolor{purple}{RGB}{204,153,255}
\definecolor{lightgray}{rgb}{.95,0.95,0.95}
\definecolor{lightgreen}{rgb}{0.3,0.5,0.3}
\definecolor{darkgreen}{rgb}{0.05,0.3,0.05}

\newcommand{\brackets}[1]{\left\{#1\right\}}
\newcommand{\bbm}[1]{{\bar{\bm #1}}}

\newcommand{\tbm}[1]{{\tilde{\bm #1}}}
\newcommand{\cbm}[1]{{\check{\bm #1}}}
\newcommand{\hbm}[1]{{\hat{\bm #1}}}
\newcommand{\inv}{^{-1}}

\newcommand{\rfield}{\mathbb{R}}

\newcommand{\tr}[1]{\mathop{\rm Tr}\left(#1\right)}

 \newcommand{\define}{\triangleq}

\newcommand{\intinfty}{\int_{-\infty}^\infty}

\newcommand{\prob}[1]{\mathop{\textrm{P}} \brackets{#1} }

\newcommand{\expected}[1]{\mathop{\textrm{E}}\brackets{#1} }

\newcommand{\pd}{P_\text{D}}

\newcommand{\st}{\mathop{\text{s.t.}}}

\newtheorem{myproposition}{Proposition}
\newtheorem{myremark}{Remark}
\newtheorem{myproblemstatement}{Problem Statement}
\newtheorem{mylemma}{Lemma}
\newtheorem{mytheorem}{Theorem}
\newtheorem{mydefinition}{Definition}
\newtheorem{mycorollary}{Corollary}

\def\myQED      {\hfill$\blacksquare$\vspace{0.3cm}}

\renewcommand{\st}{\text{s. to}}

\newcommand{\hb}[1]{\hat{\bar{#1}} }

\renewcommand{\prob}[1]{\mathop{\mathbb{P}} \brackets{#1} }
\renewcommand{\expected}[1]{\mathop{\mathbb{E}}\brackets{#1} }

\newcommand{\shadowing}{s} 

\newcommand{\srn}{M}  
\newcommand{\srind}{m}  
\newcommand{\txsgpsdsrn}[1]{\phi_{#1}}  
\newcommand{\snn}{N}  

\newcommand{\snind}{{\bm x}}  
\newcommand{\snindtt}{{\bm x}_{n_t}}  
\newcommand{\snindgen}{{\bm x'}}  
\newcommand{\cm}{\bm G} 
\newcommand{\chsrnsnn}[2]{l_{#1}(#2)}  
\newcommand{\ch}{\bm l}  
\newcommand{\chsnn}[1]{\ch({#1})}  
\newcommand{\chestsnn}[1]{\hbm l({#1})}  
\newcommand{\chsrn}[1]{l_{#1}}  
\newcommand{\wsnn}[1]{\chsnn{#1}}  
\newcommand{\w}{\bm l}  
\newcommand{\weln}[1]{\chsrn{#1}}    
\newcommand{\wsnnstn}[2]{\ch\atstn{{#2}}({#1})}
\newcommand{\rxsgpsdsnnfn}[2]{\Gamma(#1,#2)}  
\newcommand{\flesd}{G}  
\newcommand{\powsrnsnn}[2]{\Phi_{#1}(#2)} 
\newcommand{\powsnn}[1]{\bm \phi({#1})} 
\newcommand{\powsnnT}[1]{\bm \phi^T({#1})} 
\newcommand{\qensnn}[1]{\hat\pi_Q({#1})}
\newcommand{\powflsnn}[1]{\pi({#1})} 
\newcommand{\powflsnnmpsnn}[2]{\pi_{#2}(#1)} 
\newcommand{\powflestsnn}[1]{\hat \pi({#1})} 
\newcommand{\qregn}{R} 
\newcommand{\qreglim}{\tau}

\newcommand{\Q}{Q}  
\newcommand{\m}{\bm \phi}  
\newcommand{\mbound}{\bar{\varphi}}  
\newcommand{\msnn}[1]{\powsnn{#1}}  
\newcommand{\msnnT}[1]{\powsnnT{#1}}  
\newcommand{\msnnmpsnn}[2]{\bm \phi_{#2}(#1)}  
\newcommand{\msnnmpsnnT}[2]{\bm \phi^T_{#2}(#1)}  
\newcommand{\mpsnn}{P} 
\newcommand{\mpsnind}{p} 
\newcommand{\qobssnn}[1]{\qensnn{#1}}  
\newcommand{\obs}{\bm y}  
\newcommand{\obsel}{y}  
\newcommand{\obssnn}[1]{y({#1})}  
\newcommand{\obssnnmpsnn}[2]{y_{#2}(#1)}  
\newcommand{\sspace}{\mathcal{H}}

\newcommand{\cvec}{\bm c}

\newcommand{\dvec}{\bm \theta}

\newcommand{\tcvec}{\tbm c}

\newcommand{\tbRmat}{\tilde{\bbm K}}

\newcommand{\xivec}{\bm \xi}

\newcommand{\alphavec}{\bm \alpha}
\newcommand{\alphasvec}{\bm \beta}

\newcommand{\change}[1]{\textcolor{blue}{#1}}
\newcommand{\Kmat}{\bm K}
\newcommand{\Kmatsm}{\cbm K}

\newcommand{\cPhimat}{{\bm \Phi}_0}
\newcommand{\cbPhimat}{\bar{\bm \Phi}_0}
\newcommand{\kermat}{\bm K}
\newcommand{\bPhimat}{\bbm \Phi}
\newcommand{\Smat}{\bm S}
\newcommand{\instregerr}{\mathcal{C}}
\newcommand{\step}{\mu_t}

\newcommand{\stind}{t}  
\newcommand{\atstn}[1]{^{(#1)}} 
\newcommand{\stn}{T} 
\newcommand{\nterms}{I}
\newcommand{\dimn}{d}

\newcommand{\wbasiswbasisnsnn}[2]{\bm B_{#1}(#2)}
\newcommand{\wbasiswbasisnsnns}[2]{B_{#1}(#2)} 
\newcommand{\wbasisind}{{\nu}}
\newcommand{\wbasisn}{{N_B}}

\newcommand{\sspacep}{{\sspace'}}
\newcommand{\Psimat}{\bm B}
\newcommand{\Psimatsm}{\cbm B}
\newcommand{\proj}{\bm P_{\Psimat}^\perp}
\newcommand{\projz}{\bm P_{\Psimatsm}^\perp}

\newcommand{\canbasisdimnvecn}[2]{\bm e_{#1,#2}}
\newcommand{\aset}{\mathcal{A}}

\newcommand{\qenestsnnmpsnn}[2]{\hat \pi_{Q,#2}(#1)}

\usepackage{algorithmic}
\usepackage[ruled,vlined,resetcount]{algorithm2e}

\newtheorem{proposition}{Proposition}
\newtheorem{theorem}{Theorem}
\newtheorem{definition}{Definition}

\newtheorem{remark}{Remark}

\DeclareMathOperator*{\argmin}{arg\,min}

\renewcommand{\change}[1]{\textcolor{black}{#1}}

\renewcommand{\pd}{{(\bm x)}}
\renewcommand{\define}{:=}
\allowdisplaybreaks

\begin{document}

\title{Learning Power Spectrum Maps \\from Quantized Power
  Measurements}
\author{Daniel Romero, \emph{Member, IEEE},
Seung-Jun Kim, \emph{Senior Member, IEEE}, \\Georgios B. Giannakis,
\emph{Fellow, IEEE}, and Roberto L\'opez-Valcarce, \emph{Member, IEEE}
\thanks{
  D. Romero and G. B. Giannakis were supported by ARO grant
  W911NF-15-1-0492 and NSF grant 1343248. S.-J. Kim was supported by
  NSF grant 1547347. D. Romero and R. L\'opez-Valcarce were supported
  by the Spanish Ministry of Economy and Competitiveness and the
  European Regional Development Fund (ERDF) (projects
  TEC2013-47020-C2-1-R, TEC2016-76409-C2-2-R and TEC2015-69648-REDC),
  and by the Galician Government and ERDF (projects GRC2013/009,
  R2014/037 and ED431G/04).

  D. Romero is with the Dept. of Information and Communication
  Technology, Univ. of Agder, Norway.  S.-J. Kim is with the Dept. of
  Computer Sci. \& Electrical Eng., Univ. of Maryland, Baltimore
  County, USA. D. Romero was and G. B. Giannakis is with the Dept. of
  ECE and Digital Tech. Center, Univ. of Minnesota, USA.  D. Romero
  was and R. L\'opez-Valcarce is with the Dept. of Signal Theory and
  Communications, Univ. of Vigo, Spain. E-mails: daniel.romero@uia.no,
  sjkim@umbc.edu, georgios@umn.edu, valcarce@gts.uvigo.es.

  Parts of this work have been presented at the IEEE International
  Conference on Acoustics, Speech, and Signal Processing, Brisbane
  (Australia), 2015; at the Conference on Information Sciences and
  Systems, Baltimore (Maryland), 2015; and at the IEEE International
  Workshop on Computational Advances in Multi-sensor Adaptive
  Processing, Canc\'un (M\'exico), 2015.}  }

\maketitle

\begin{abstract}
  Power spectral density (PSD) maps providing the distribution of RF
  power across space and frequency are constructed using power
  measurements collected by a network of low-cost sensors. By
  introducing linear compression and quantization to a small number of
  bits, sensor measurements can be communicated to the fusion center
  with minimal bandwidth requirements. Strengths of data- and
  model-driven approaches are combined to develop estimators capable
  of incorporating multiple forms of spectral and propagation prior
  information while fitting the rapid variations of shadow fading
  across space. To this end, novel nonparametric and semiparametric
  formulations are investigated.  It is shown that PSD maps can be
  obtained using support vector machine-type solvers. In addition to
  batch approaches, an online algorithm attuned to real-time operation
  is developed.  Numerical tests assess the performance of the novel
  algorithms.
\end{abstract}

\section{Introduction} 

Power spectral density (PSD) cartography relies on sensor measurements
to map the radiofrequency (RF) signal power distribution over a
geographical region. The resulting maps are instrumental for various
wireless network management tasks, such as power control, interference
mitigation, and
planning~\cite{kim2011kriged,dallanese2011powercontrol}.  { For
  instance, PSD maps benefit wireless network planning by revealing
  the location of crowded regions and areas of weak coverage, thereby
  suggesting where new base stations should be deployed. Because they
  characterize how RF power distributes per channel across space, PSD
  maps are also useful  to increase frequency reuse and mitigate
  interference in cellular systems. In addition, PSD maps enable
  opportunistic transmission in cognitive radio systems by unveiling
  underutilized ``white spaces'' in time, space, and
  frequency~\cite{axell2012tutorial,yucek2009surveysensing}.
  Different from conventional spectrum sensing techniques, which
  assume a common spectrum occupancy over the entire sensed
  region~\cite{quan2008collaborativewideband,ariananda2014cooperative,mehanna2013frugal},
  PSD cartography accounts for variability across space and therefore
  enables a more aggressive spatial reuse.  }

  A number of interpolation and regression techniques have been
  applied to construct RF power maps from power measurements. Examples
  include kriging~\cite{alayafeki2008cartography}, compressive
  sampling~\cite{jayawickrama2013compressive}, dictionary
  learning~\cite{kim2011link,kim2013dictionary}, sparse Bayesian
  learning~\cite{huang2014sparsebayesian}, and matrix
  completion~\cite{ding2015devicetodevice}.  These maps describe how
  power distributes over space but not over frequency. To accommodate
  variability along the frequency dimension as well, a basis expansion
  model was introduced
  in~\cite{bazerque2010sparsity,dallanese2012gslasso,bazerque2011splines}
  to create {PSD maps} from periodograms. To alleviate the high power
  consumption and bandwidth needs that stem from obtaining and
  communicating these periodograms,~\cite{mehanna2013frugal} proposed
  a low-overhead sensing scheme based on single-bit data along the
  lines of~\cite{argg2006tsp}. However, this scheme assumes that
  the PSD is constant across space.

To summarize, existing spectrum cartography approaches either
construct \emph{power} maps from \emph{power} measurements, or,
\emph{PSD} maps from \emph{PSD} measurements. In contrast, the main
contribution of this paper is to present algorithms capable of
estimating \emph{PSD} maps from \emph{power} measurements, thus
attaining a more efficient extraction of the information contained in
the observations than existing methods.  Therefore, the proposed
approach enables the estimation of the RF power distribution over
\emph{frequency and space} using low-cost low-power sensors since only
power measurements are required.

To facilitate practical implementations with sensor networks, where
the communication bandwidth is limited, overhead is reduced by
adopting two measures. First, sensor measurements are quantized to
a small number of bits. Second, the available prior information is
efficiently captured in both frequency and spatial domains, thus
affording strong quantization while minimally sacrificing the quality
of map estimates.

Specifically, a great deal of frequency-domain prior information about
impinging communication waveforms can be collected from spectrum
regulations and standards, which specify bandwidth, carrier
frequencies, transmission masks, roll-off factors, number of
sub-carriers, and so
forth~\cite{vazquez2011guardbands,romero2013wideband}. To exploit
this information, a basis expansion model is adopted, which allows the
estimation of the power of each sub-channel and background noise as a
byproduct. The resulting estimates can be employed to construct
signal-to-noise ratio (SNR) maps, which reveal weak coverage areas,
and alleviate the well-known noise uncertainty problem in cognitive
radio~\cite{tandra2008snrwalls}.

To incorporate varying degrees of prior information in the spatial
domain, nonparametric and semiparametric estimators are developed
within the framework of kernel-based learning for vector-valued
functions~\cite{micchelli2005vectorvalued}. Nonparametric estimates
are well suited for scenarios where no prior knowledge about the
propagation environment is available due to their ability to
approximate any spatial field with arbitrarily high
accuracy~\cite{carmeli2010vector}. In many cases, however, one may
approximately know the transmitter locations, the path loss exponent,
or even the locations of obstacles or reflectors. The proposed
semiparametric estimators capture these forms of prior information
through a basis expansion in the spatial domain.

Although the proposed estimators can be efficiently implemented in
batch mode, limited computational resources may hinder real-time
operation if an extensive set of measurements is to be processed.
This issue is mitigated here through an online nonparametric
estimation algorithm based on stochastic gradient descent. Remarkably,
the proposed algorithm can also be applied to (vector-valued) function
estimation setups besides spectrum cartography.
  
The present paper also contains two theoretical contributions to
machine learning. First, a neat connection is established between
robust (possibly vector-valued) function estimation from quantized
measurements and support vector
machines~(SVMs)~\cite{scholkopf2001,smola2004tutorial,smola1998connection}.
Through this link, theoretical guarantees and efficient
implementations of SVMs permeate to the proposed methods.  Second,
the theory of kernel-based learning for vector-valued functions is
extended to accommodate semiparametric estimation.  The resulting
methods are of independent interest since they can be applied
beyond the present spectrum cartography context and subsume, as a
special case, thin-plate splines
regression~\cite{wang2001deformation,bazerque2011splines}.

The rest of the paper is organized as follows. Sec.~\ref{sec:om}
presents the system model and formulates the
problem. Sec.~\ref{sec:cl} proposes nonparametric and semiparametric
PSD map estimation algorithms operating in batch mode, whereas
Sec.~\ref{sec:oi} develops an online solver. Finally,
Sec.~\ref{sec:sim} presents some numerical tests and
Sec.~\ref{sec:con} concludes the paper with closing remarks and
research directions.

\noindent {\bf Notation.} The cardinality of set $\aset$ is denoted by
$|\aset|$. Scalars are denoted by lowercase letters, column vectors by
boldface lowercase letters, and matrices by boldface uppercase
letters. Superscript $^T$ stands for transposition, and $^H$ for
conjugate transposition. The $(i,j)$th entry ($j$th column) of matrix
$\bm A$ is denoted by $a_{i,j}$ ($\bm a_j$). The Kronecker product is
represented by the symbol $\otimes$. The Khatri-Rao product is defined
for $\bm A:=[\bm a_1,\bm a_2,\ldots,\bm a_{N}] \in \mathbb{C}^{M_1
  \times N}$ and $\bm B:=[\bm b_1,\bm b_2,\ldots,\bm b_N] \in
\mathbb{C}^{M_2 \times N}$ as $\bm A \odot \bm B \define [\bm a_1
  \otimes \bm b_1,\ldots,\bm a_N \otimes \bm b_N] \in \mathbb{C}^{M_1
  M_2 \times N}$.  The entrywise or Hadamard product is defined as
$(\bm A\circ \bm B)_{i,j}\define a_{i,j}b_{i,j}$. Vector
$\canbasisdimnvecn{M}{m}$ is the $m$-th column of the $M\times M$
identity matrix $\bm{I}_M$, whereas $\bm 0_M$ and $\bm 1_M$ are the
vectors of dimension $M$ with all zeros and ones, respectively. Symbol
$\expected{\cdot}$ denotes expectation, $\prob{\cdot}$ probability,
$\tr{\cdot}$ trace, $\lambda_{\max}(\cdot)$ largest eigenvalue, and
$\star$ convolution. Notation $\lceil \rho \rceil$ (alternatively
$\lfloor \rho \rfloor$) represents the smallest (largest) integer $z$
satisfying $z\geq \rho$ ($z\leq \rho$).

\section{System Model and Problem Statement}
\label{sec:om}

Consider $\srn-1$ transmitters located in a geographical region
$\mathcal{R} \subset \rfield^\dimn$, where $\dimn$ is
typically\footnote{One may set $\dimn=1$ for maps along roads or
  railways, or even $\dimn=3$ for applications involving aerial
  vehicles or urban environments.} $2$.  Let $\txsgpsdsrn{\srind}(f)$
denote the transmit-PSD of the $m$-th transmitter and let
$\chsrnsnn{\srind}{\snind}$ represent the gain of the channel between
the $m$-th transmitter and location $\snind\in \mathcal{R}$, which is
assumed frequency flat to simplify the exposition; see
Remark~\ref{remark:frequencyselective}. If the $\srn-1$ transmitted
signals are uncorrelated, the PSD at location $\snind$ is given by
\begin{align}
\label{eq:rxsgpsdsnn}
\rxsgpsdsnnfn{\snind}{f}
=\sum_{\srind=1}^{\srn}\chsrnsnn{\srind}{\snind}
\txsgpsdsrn{\srind}(f)
\end{align}
where $\chsrnsnn{\srn}{\snind}$ is the noise power and
$\txsgpsdsrn{\srn}(f)$ is the noise PSD, normalized to
$\intinfty\txsgpsdsrn{\srn}(f)df=1$. In view of \eqref{eq:rxsgpsdsnn},
one can also normalize $\txsgpsdsrn{\srind}(f)$,
$\srind=1,\ldots,\srn-1$, without any loss of generality to satisfy
$\intinfty\txsgpsdsrn{\srind}(f)df=1$ by absorbing any scaling factor
into $\chsrnsnn{\srind}{\snind}$.  Since such a scaling factor equals
the transmit power, the coefficient $\chsrnsnn{\srind}{\snind}$
actually represents the power of the $\srind$-th signal at location
$\snind$.

Often in practice, the normalized PSDs
$\{\txsgpsdsrn{\srind}(f)\}_{m=1}^{M-1}$ are approximately known since
transmitters typically adhere to publicly available standards and
regulations, which prescribe spectral masks, bandwidths, carrier
frequencies, roll-off factors, number of pilots, and so
on~\cite{vazquez2011guardbands,romero2013wideband}.  If unknown, the
methods here carry over after setting
$\{\txsgpsdsrn{\srind}(f)\}_{m=1}^{M-1}$ to be  a
frequency-domain basis expansion
model~\cite{bazerque2010sparsity,bazerque2011splines}. For this
reason, the rest of the paper assumes that
$\{\txsgpsdsrn{\srind}(f)\}_{m=1}^{M}$ are known.

\begin{figure}[t]
\begin{minipage}[b]{\linewidth}
\center
\includegraphics[height=1.5cm]{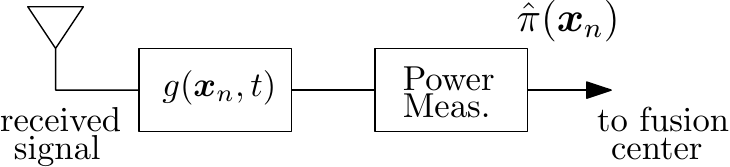}
\end{minipage}
~\\
\begin{minipage}[b]{\linewidth}
\center
\includegraphics[height=1.5cm]{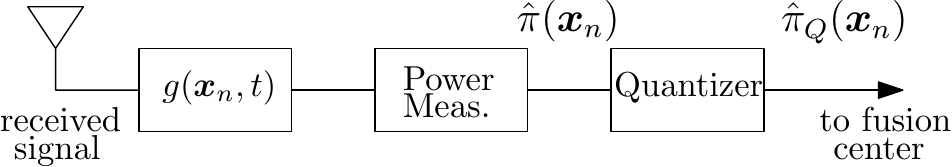}
\end{minipage}
\caption{{\color{black}Sensor architectures without (top) and with
  quantization  (bottom).}}
\label{fig:sensors}
\end{figure}


The goal is to estimate the PSD map $\rxsgpsdsnnfn{\snind}{f}$ from
the measurements gathered by $N$ sensors with locations
$\{\snind_n\}_{n=1}^N\subset \mathcal{R}$. In view of
\eqref{eq:rxsgpsdsnn}, this task is tantamount to estimating
$\{\chsrnsnn{\srind}{\snind}\}_{m=1}^{M}$ at every spatial coordinate
$\snind$.

To minimize hardware costs and power consumption, this paper adopts
the sensor architecture in~Fig.~\ref{fig:sensors}.  The
\emph{ensemble} power at the output of the filter of the $n$-th sensor
is $\pi (\snind_n) =
\intinfty|\flesd(\snind_n,f)|^2\rxsgpsdsnnfn{\snind_n}{f}df$, where
$G(\snind_n,f)$ denotes the frequency response of the receive
filter. From \eqref{eq:rxsgpsdsnn}, it follows that
\begin{align}
\label{eq:lmm}
\powflsnn{\snind_n}
=\sum_{\srind=1}^{\srn}\chsrnsnn{\srind}{\snind_n} \powsrnsnn{\srind}{\snind_n}
= {\bm \phi}^T (\snind_n) \chsnn{\snind_n}
\end{align}
where $\powsrnsnn{\srind}{\snind_n} :=\intinfty|\flesd(\snind_n,f)|^2
\txsgpsdsrn{\srind}(f)df$ can be thought of as the contribution of the
$m$-th transmitter per unit of $\chsrnsnn{\srind}{\snind_{n}}$ to
the power at the output of the receive filter of the $n$-th sensor;
whereas $\powsnn{\snind_n}:=$
$[\powsrnsnn{1}{\snind_n},\ldots,\powsrnsnn{\srn}{\snind_n}]^T$ and
$\chsnn{\snind_n} := [\chsrnsnn{1}{\snind_n},\ldots ,$
  $\chsrnsnn{\srn}{\snind_n}]^T$ are introduced to simplify notation.
The $n$-th sensor obtains an estimate $\powflestsnn{\snind_n}$ of
$\powflsnn{\snind_n}$ by measuring the signal power at the output of
the filter over a certain time window. In general, 
$\powflestsnn{\snind_n} \neq \powflsnn{\snind_n} $ due to the finite
length of this observation window.

Different from the measurement model
in~\cite{bazerque2010sparsity,dallanese2012gslasso,bazerque2011splines},
where sensors  obtain and communicate periodograms, the
proposed scheme solely involves power measurements, thereby reducing
 sensor costs and bandwidth requirements.  To further reduce
bandwidth, $\powflestsnn{\snind_n}$ can be quantized as illustrated in
the bottom part of Fig.~\ref{fig:sensors}. When uniform quantization is
used,  the sensors obtain
\begin{align}
\qensnn{\snind_n} := \Q(\hat \pi(\snind_n)) := \lfloor \hat \pi(\snind_n)/(2\epsilon)\rfloor \;,\;\; n=1,\ldots,N\label{eq:quant}
\end{align}
where $2\epsilon$ is the quantization step; see also
Remark~\ref{remark:nonuniform}.  If $\qregn$ denotes the number of
quantization levels, which depends on $\epsilon$ and the range of
$\hat \pi(\snind_n)$, the number of bits needed is now just $\lceil
\log_2 \qregn\rceil$.  Depending on how accurate $\hat \pi(\snind_n)$
is, either $\Q(\powflsnn{\snind_n})=\Q(\powflestsnn{\snind_n})$ or
$\Q(\powflsnn{\snind_n})\neq\Q(\powflestsnn{\snind_n})$. The latter
event is termed \emph{measurement error} and is due to the
finite length of the aforementioned time window.

Finally, the sensors communicate the measurements $\{\hat
\pi({\snind_n})\}_{n=1}^N$ or $\{\qensnn{\snind_n}\}_{n=1}^N$ to the
fusion center. Given these measurements, together with
$\{\txsgpsdsrn{\srind}(f)\}_{m=1}^{M}$ and $\{\bm x_{n}\}_{n=1}^{N}$,
the \emph{problem} is to estimate
$\{\chsrnsnn{\srind}{\snind}\}_{m=1}^{M}$ at
every~$\snind\in\mathcal{R}$.  The latter can be viewed individually
as $M$ functions of the spatial coordinate $\bm x$, or, altogether as
a vector field $\bm l:\rfield^d \rightarrow \rfield^M$, where $\bm
l(\bm x)\define [l_1(\bm x),\ldots,l_M(\bm x)]^T$. Thus, estimating
the PSD map in \eqref{eq:rxsgpsdsnn} is in fact a problem of
estimating a vector-valued function.


\noindent

{
\begin{remark}[Frequency-selective channels]
\label{remark:frequencyselective}
 If the channels are not frequency-flat, then each term in the sum of
 \eqref{eq:rxsgpsdsnn} can be decomposed into multiple components of
 smaller bandwidth in such a way that each one \emph{sees} an approximately
 frequency-flat channel. To ensure that these components are
 uncorrelated, one can choose their frequency supports to be disjoint.
\end{remark}
}

\begin{remark}Sensors can also operate digitally. In this case, one
  could implement the receive filters to have pseudo-random impulse
  responses~\cite{mehanna2013frugal}. Selecting distinct seeds for the
  random number generators of different sensors yields linearly
  independent $\{\powsnn{\snind_n}\}_{n=1}^N$ with a high probability,
  which ensures identifiability of
  $\{\chsrnsnn{\srind}{\snind_n}\}_{m=1}^{M}$~(cf.~\eqref{eq:lmm}).
\end{remark}

\noindent
\begin{remark}
\label{remark:a2ic}
 If a wideband map is to be constructed, then Nyquist-rate sampling
 may be too demanding for low-cost sensors. In this scenario, one can
 replace the filter in Fig.~\ref{fig:sensors} with an
 analog-to-information-converter
 (A2IC)~\cite{tropp2010beyond,romero2013wideband}. To see that
 \eqref{eq:lmm} still holds and therefore the proposed schemes still
 apply, let $\cm({\snind_n})$ represent the compression matrix of the
 $n$th sensor, which multiplies raw measurement blocks to yield
 compressed data blocks~\cite{romero2013wideband}. The ensemble power
 of the latter is proportional to $\pi (\bm x_n) \define
 \tr{\cm({\snind_n}) \bm \Sigma({\snind_n}) \cm^T({\snind_n})}$, where
 $\bm \Sigma({\snind_n}) = \sum_{\srind=1}^\srn
 \chsrnsnn{\srind}{\snind_n} \bm \Sigma_\srind$ denotes the covariance
 matrix of the uncompressed data blocks, and $\bm \Sigma_\srind$ the
 covariance matrix of the blocks transmitted from the $m$-th
 transmitter. Combining both equalities yields \eqref{eq:lmm} upon
 defining $\powsrnsnn{\srind}{\snind_n} := \tr{ \cm ({\snind_n}) \bm
   \Sigma_\srind \cm^T ({\snind_n})}$.
\end{remark}

\section{Learning PSD Maps}
\label{sec:cl}

This section develops various PSD map estimators offering different
bandwidth-performance trade-offs.  First, Sec.~\ref{sec:nprnq} puts
forth a nonparametric estimator to recover $\bm l(\bm x):=[l_1(\bm
  x),\ldots, l_M(\bm x)]^T$ from un-quantized power measurements. To reduce
bandwidth requirements, this approach is extended in
Sec.~\ref{sec:npr} to accommodate quantized data. The detrimental
impact of strong quantization on the quality of map estimates is
counteracted in Sec.~\ref{sec:spr} by leveraging propagation prior
information. For simplicity, these methods are presented for the
scenario where each sensor obtains a single measurement, whereas
general versions accommodating multiple measurements per sensor are
outlined in Sec.~\ref{sec:emms}.

\subsection{{Estimation via nonparametric  regression}}
\label{sec:nprnq}

This section reviews the background in kernel-based learning necessary
to develop the cartography tools in the rest of the paper and presents
an estimator to obtain PSD maps from un-quantized measurements.

Kernel-based regression seeks estimates among a wide class of
functions termed \emph{reproducing kernel Hilbert space} (RKHS). In
the present setting of vector-valued functions, such an RKHS is given
by ${\cal H}:=\{ \ch (\snind) = \sum_{n=1}^\infty \kermat(\bm x,\tbm
x_n) \tbm c_n :\tbm x_n \in \rfield^d, \tbm c_n \in
\rfield^M\}$~\cite{micchelli2005vectorvalued,carmeli2010vector}, where
$\{\tbm c_n\}_{n=1}^\infty$ are $M\times1$ expansion coefficient
vectors and $\bm K (\bm x,\bm x')$ is termed reproducing kernel
map. The latter is any matrix-valued function $\bm K (\bm x,\bm
x'):\rfield^d \times \rfield^d \rightarrow \rfield^{M\times M}$ that
is (i) symmetric, meaning that $\bm K(\bm x,\bm x')=\bm K(\bm x',\bm
x)$ for any $\bm x$ and $\bm x'$; and (ii) positive (semi)definite,
meaning that the square matrix having $\bm K (\tbm x_n,\tbm x_{n'})$
as its $(n,n')$ block is positive semi-definite for any $\{\tbm
x_1,\ldots,\tbm x_{\tilde N}\}$. Remark~\ref{remark:kernel} guides the
selection of functions $\bm K (\bm x,\bm x')$ qualifying as
reproducing kernels.

As any Hilbert space, an RKHS has an associated inner product, not
necessarily equal to the classical $\langle \bm f, \bm g \rangle = \int
\bm f^T(\bm x)\bm g(\bm x)d\bm x$. Specifically, the inner product between two RKHS
functions $ \ch (\snind) \define \sum_{n=1}^{\tilde N} \kermat(\bm x,\tbm
x_n) \tbm c_n$ and $ \ch' (\snind) \define \sum_{n=1}^{\tilde N'}
\kermat(\bm x,\tbm x_n') \tbm c_n'$ can be obtained through the
reproducing kernel as
\begin{align}
\label{eq:repprop}
  \langle \ch, \ch' \rangle_\mathcal{H}= 
\sum_{n=1}^{\tilde N} \sum_{n'=1}^{\tilde N'} \tbm c^T_n\bm K
(\tbm x_n,\tbm x_{n'}') \tbm c_{n'}'.
\end{align}
This expression is referred to as the \emph{reproducing property} and
is of paramount importance since it allows the computation of function
inner products without integration. From \eqref{eq:repprop}, the
induced RKHS norm of $\ch$ can be written as $\| \ch \|_{\cal H}^2
:=\langle \ch,\ch\rangle=\sum_{n=1}^\infty \sum_{n'=1}^\infty \tbm
c^T_n\bm K (\tbm x_n,\tbm x_{n'}) \tbm c_{n'}$ and is widely used as a
proxy for smoothness of $\ch$.

Kernel-based methods confine their search of estimates to functions in
$\mathcal{H}$, which is not a limitation since an extensive class of
functions, including any continuous function vanishing at infinity,
can be approximated with arbitrary accuracy by a function in
$\mathcal{H}$ for a properly selected
kernel~\cite{carmeli2010vector}. However, function estimation is
challenging since (i) any finite set of samples generally admits
infinitely many interpolating functions in $\mathcal{H}$ and (ii) the
estimate $\hat \ch$ does not generally approach the estimated function
$\ch$ even for an arbitrary large number of noisy samples if the
latter are overfitted.  To mitigate both issues, kernel regression
seeks estimates minimizing the sum of two terms, where the first
penalizes estimates deviating from the observations and the second
promotes smoothness.  Specifically, if ${\cal L}(e_n)$ denotes a loss
function of the measurement error $e_n := \hat \pi(\snind_n) - {\bm
  \phi}^T (\snind_n) \chsnn{\snind_n}$, the nonparametric kernel-based
regression estimate of $\ch$ is~\cite{berlinet2004}
\begin{equation}
\hat \ch \define  \argmin_{\ch \in \cal H} \frac{1}{N}\sum_{n=1}^N
     {\cal L}\left( \hat \pi (\snind_n)- {\bm \phi}^T (\snind_n) \chsnn{\snind_n}\right) +\lambda \| \ch \|_{\cal H}^2\label{eq4}
\end{equation}
where the user-selected scalar $\lambda>0$ controls the tradeoff
between  fitting the data and  smoothness of the
estimate, captured by its RKHS norm.

Since $\cal H$ is infinite-dimensional, solving~\eqref{eq4} directly
is generally not possible with a finite number of
operations. Fortunately, the so-called \emph{representer theorem} (see
e.g.,~\cite{scholkopf2001,argyriou2014unifying}) asserts that $\hat
\ch$ in \eqref{eq4} is of the form
\begin{equation}\label{eq5}
\sum_{n=1}^N \bm K (\bm x,\bm x_{n}) \bm c_{n} := \bm K (\bm x)\bm c
\end{equation}
for some $\{\bm c_{n}\}_{n=1}^N$, where $\bm K (\bm x):= [\bm K (\bm
  x,\bm x_1) ,\ldots, \bm K (\bm x,\bm x_N)]$ is of size $M\times MN$,
and $\bm c := [\bm c_1^T ,\ldots, \bm c_N^T]^T $ is $MN\times 1$. In
words, the solution to \eqref{eq4} admits a kernel expansion around
the sensor locations $\{\bm x_n\}_{n=1}^N$. From \eqref{eq5}, it
follows that finding $\hat \ch$ amounts to finding $\bm c$, but the
latter can easily be obtained by solving the problem that results from
substituting \eqref{eq5} into~\eqref{eq4}:
\begin{equation}\label{eq6p}
\hat{\bm c} \define \argmin_{\bm c} \sum_{n=1}^N {\cal L}( \hat \pi (\snind_n)- {\bm \phi}^T (\bm x_n)  \bm K (\bm x_n)\bm c) +\lambda N \bm c^T \bm K \bm c.
\end{equation}
In \eqref{eq6p}, the $MN\times MN$ matrix $\bm K$ is formed to have
$\bm K (\bm x_n,\bm x_{n'})$ as its $(n,n')$-th block.  Clearly, the
minimizer of \eqref{eq4} can then be recovered as $\hat \ch (\bm x)=
\bm K (\bm x) \hat {\bm c}$.

The loss function $\mathcal{L}$ is typically chosen to be convex.  The
simplest example is the squared loss $\mathcal{L}_2(e_n):=e_n^2$, with
the resulting $\hat \ch$ referred to as the \emph{kernel ridge
  regression} estimate. Defining $\hbm \pi:= [\hat \pi(\snind_1)
  ,\ldots, \hat\pi(\snind_N)]^T$, $\bm \Phi := [\bm \phi (\bm
  x_1),\ldots, \bm \phi (\bm x_N)]$, and $\bm \Phi_0 := \bm I_N \odot
\bm \Phi$, expression \eqref{eq6p} becomes
\begin{align}
\nonumber
\hat{\bm c} &= \argmin_{\bm c\in \rfield^{MN}}  || \hbm \pi- {\bm
  \Phi_0^T}   \bm K\bm c||^2_2 +\lambda N \bm c^T \bm K \bm c
\\
&= (\bm \Phi_0 \bm \Phi_0^T \bm K+\lambda N\bm I_{MN})\inv
\bm \Phi_0 \hbm \pi\label{eq:optcridgeregression}.
\end{align}

Besides its simplicity of implementation, the estimate $\hat \ch (\bm
x)= \bm K (\bm x) \hat {\bm c}$ with $\hat{\bm c}$ as in
\eqref{eq:optcridgeregression} offers a twofold advantage over
existing cartography schemes. First, existing estimators relying on
\emph{power measurements} can only construct~\emph{power
  maps}~\cite{alayafeki2008cartography,jayawickrama2013compressive,huang2014sparsebayesian,kim2011link,kim2013dictionary},
whereas the proposed method is capable of obtaining \emph{PSD maps}
from the same measurements. On the other hand, existing methods for
estimating \emph{PSD maps} require \emph{PSD measurements}, that is,
every sensor must obtain and transmit periodograms to the fusion
center. This necessitates a higher communication bandwidth, longer sensing
time, and more costly sensors than required
here~\cite{bazerque2010sparsity,dallanese2012gslasso,bazerque2011splines}.


\begin{remark}
\label{remark:kernel}
The choice of the kernel considerably affects the estimation
performance when the number of observations is small. Thus, it is
important to choose a kernel that is well-suited to the spatial
variability of the true $\bm l$. To do so, one may rely on cross
validation, historical data~\cite[Sec.~2.3]{scholkopf2001}, or
multi-kernel approaches~\cite{jbgg2013spmag}.  Although specifying
matrix-valued kernels is more challenging than specifying their scalar
counterparts ($M=1$)~\cite{micchelli2005vectorvalued}, a simple but
effective possibility is to construct a diagonal kernel as $\bm K (
\bm x , \bm x') = {\rm diag} (k_1(\bm x,\bm x'), \ldots ,k_M (\bm x ,
\bm x') )$ where $\{k_m (\bm x , \bm x')\}_{m=1}^M$ are valid scalar
kernels.  For example, $k_m (\bm x , \bm x')$ can be the popular
Gaussian kernel $k_m (\bm x,\bm x') = \exp (-\| \bm x - \bm
x'\|^2/\sigma_m^2)$, where $\sigma_m^2>0$ is user selected.
\end{remark}

\subsection{Nonparametric regression from quantized data}
\label{sec:npr}

The scheme in Sec.~\ref{sec:nprnq} offers a drastic bandwidth
reduction relative to competing PSD map estimators since only
scalar-valued measurements need to be communicated to the fusion
center.  The methods in this section accomplish a further reduction by
accommodating quantized measurements.

Recall from Sec.~\ref{sec:om}, that $\{\qensnn{\snind_n}\}_{n=1}^N$
denote the result of uniformly quantizing the power measurements
$\{\hat \pi({\snind_n})\}_{n=1}^N$; see also
Remark~\ref{remark:nonuniform}. The former essentially convey
\emph{interval information} about the latter, since \eqref{eq:quant}
implies that $\hat \pi(\bm x_n)$ is contained in the interval $[y(\bm
  x_n)-\epsilon, y(\bm x_n)+\epsilon)$, where
  $\obssnn{\snind_n}:=[2\qobssnn{\snind_n}+1]\epsilon$. Note that
  $\obssnn{\snind_n}$ is in fact the centroid of the
  $\qobssnn{\snind_n}$-th quantization interval.

 To account for the uncertainty within such an interval, one can
replace $\hat \pi (\snind_n)$ in 
 \eqref{eq4} with $y (\snind_n)$ as
\begin{equation}
\hat \ch = \argmin_{\ch \in \cal H} \frac{1}{N}\sum_{n=1}^N
     {\cal L}( y (\snind_n)- {\bm \phi}^T (\snind_n) \chsnn{\snind_n})
     +\lambda \| \ch \|_{\cal H}^2
\label{eq4b}
\end{equation}
and set $\mathcal{L}$ to assign no cost across all candidate functions
$\bm l$ that lead to values of $ \bm \phi^T (\bm x_n) \bm l (\bm x_n)
= \pi(\bm x_n)$ falling $\pm \epsilon$ around $y(\bm x_n)$. In other
words, such an $\mathcal{L}$ only penalizes functions $\bm l$ for
which $e_n=y(\bm x_n) - \bm \phi^T (\bm x_n) \bm l(\bm x_n)$ falls
outside of $[-\epsilon,\epsilon)$. Examples of these
  $\epsilon$-\emph{insensitive loss functions} include ${\cal
    L}_{1\epsilon} (e_n):= \max
  (0,|e_n|-\epsilon)$~\cite{scholkopf2001,smola2004tutorial}, and the
  less known ${\cal L}_{2\epsilon} (e_n):=\max
  (0,e_n^2-\epsilon)$. Incidentally, these functions endow the
  proposed estimators with robustness to outliers and promote sparsity
  in $\{e_n\}_{n=1}^N$, which is   a particularly well-motivated
  property when the number of measurement errors is small relative to
  $N$, that is, when
  $\Q(\powflsnn{\snind_n})=\Q(\powflestsnn{\snind_n})$ for \emph{most}
  values of~$n$.

The rest of this section develops solvers for \eqref{eq4b} and
establishes a link between \eqref{eq4b} and SVMs. To this end, note
that application of the representer theorem to \eqref{eq4b} yields, as
in Sec.~\ref{sec:nprnq}, an estimate $\hat \ch (\bm x)= \bm K (\bm x)
\hat {\bm c}$ with
\begin{equation}\label{eq6}
\hat{\bm c} =  \argmin_{\bm c} \sum_{n=1}^N {\cal L}( y (\snind_n)- {\bm \phi}^T (\bm x_n)  \bm K (\bm x_n)\bm c) +\lambda N \bm c^T \bm K \bm c.
\end{equation}
Now focus on ${\cal L}_{1\epsilon}$ and note that ${\cal
  L}_{1\epsilon} (e_n)=\xi_n + \zeta_n$, where $\xi_n := \max
(0,e_n-\epsilon)$ and $\zeta_n:= \max (0,-e_n - \epsilon)$
respectively quantify positive deviations of $e_n$ with respect to the
right and left endpoints of $[-\epsilon, \epsilon)$.  This implies
  that $\xi_n$ satisfies $\xi_n \geq e_n - \epsilon$ and $\xi_n \geq
  0$, whereas $\zeta_n$ satisfies $\zeta_n \geq -e_n - \epsilon$ and
  $\zeta_n \geq 0$, thus establishing the following result.
\begin{proposition} \label{prop:repth}
The problem in \eqref{eq6} with $\mathcal{L}$ the
$\epsilon$-insensitive loss function ${\cal L}_{1\epsilon}$ can be
expressed as
\begin{align}
(\hat{\bm c},\hat{\bm \xi},& \hat{\bm \zeta}) =
  \argmin_{\bm c,\bm \xi, \bm \zeta} \sum_{n=1}^N (\xi_n +\zeta_n)
+\lambda N \bm c^T \bm K \bm c   \label{eq9}\\
{\rm s. to}~~ \xi_n & \geq y (\snind_n)- {\bm \phi}^T (\bm x_n)  \bm K (\bm x_n)\bm c - \epsilon\:,\;\;\xi_n \geq 0,\; \nonumber\\
 \zeta_n & \geq - y (\snind_n) + {\bm \phi}^T (\bm x_n)  \bm K (\bm
 x_n)\bm c - \epsilon,\:\zeta_n \geq 0,\; \nonumber
\\&~~~n=1,\ldots,N \nonumber
\end{align}
where $\bm \xi\define[\xi_1,\ldots,\xi_N]^T$ and $\bm
\zeta\define[\zeta_1,\ldots,\zeta_N]^T$.
\end{proposition}
Problem \eqref{eq9} is a convex quadratic program with slack variables
$\{\xi_n,\zeta_n\}_{n=1}^N$.  Although one can obtain $(\hat{\bm
  c},\hat{\bm \xi}, \hat{\bm \zeta})$ using, for example, an
off-the-shelf interior-point solver, it will be shown that a more
efficient approach is to solve the dual-domain version of \eqref{eq9}.

To the best of our knowledge, \eqref{eq9} constitutes the first
application of an $\epsilon$-insensitive loss to estimating functions
from quantized data.  As expected from the choice of loss function and
regularizer, \eqref{eq9} is an SVM-\emph{type} problem. However,
different from existing SVMs, for which data comprises
\emph{vector-valued} noisy versions of $\{\ch(\bm
x_n)\}_{n=1}^N$~\cite[Examples 1 and 2]{micchelli2005vectorvalued},
the estimate in \eqref{eq9} relies on noisy versions of the
\emph{scalars} $\{\bm \phi^T (\bm x_n)\ch(\bm
x_n)\}_{n=1}^N$. Therefore, \eqref{eq9} constitutes a new class of SVM
for vector-valued function estimation.  As a desirable consequence of
this connection, the proposed estimate inherits the well-documented
generalization performance of existing
SVMs~\cite{micchelli2005vectorvalued,smola1998connection,carmeli2010vector}. However,
it is prudent to highlight one additional notable difference between
\eqref{eq9} and conventional SVMs that pertains to the present context
of function estimation from quantized data: whereas in the present
setting $\epsilon$ is determined by the quantization interval length,
this parameter must be delicately tuned in conventional SVMs to
control generalization performance.

The proposed estimator is \emph{nonparametric} since the number of
unknowns in \eqref{eq9} depends on the number of observations
$N$. Although this number of unknowns also grows with $M$, it is shown
next that this is not the case in the dual formulation.  To see this,
let $\bm K_0 := \bm \Phi_0^T \bm K \bm \Phi_0$ as well as $\bm y
:=[y(\bm x_1),\ldots, y(\bm x_N)]^T$. With $\bm \alpha$ and $\bm \beta$
representing the Langrange multipliers associated with the $\{\xi_n\}$
and the $\{\zeta_n\}$ constraints, the dual of \eqref{eq9} can be
easily shown to be
\begin{align}
\label{eq:probnpd}
\begin{aligned}
&(\hat{\bm \alpha},\hat{\bm \beta}) := \argmin_{\bm \alpha,\bm
    \beta\in \rfield^N}
&&\frac{1}{4\snn\lambda}(\alphavec-\bm \beta)^T \bm K_0 (\alphavec-\bm \beta )\\
&&&-(\obs-\epsilon\bm 1_\snn)^T\alphavec +(\obs+\epsilon \bm 1_\snn)^T\bm \beta\\
&\hspace{1.5cm}\st&&  \bm 0_\snn\leq \alphavec\leq \bm 1_\snn,
 \bm 0_\snn\leq \bm \beta \leq \bm 1_\snn.
\end{aligned}
\end{align}
From the Karush-Kuhn-Tucker (KKT) conditions, the primal
variables can be recovered from the dual ones using
\begin{subequations}
\label{eq:cvecoptall}
\begin{align}
\label{eq:cvecopt}
\hat \cvec &= \frac{1}{2\lambda\snn} \bm \Phi_0(\hat \alphavec- \hat {\bm  \beta} )\\
\hat \xivec &= \max(\bm 0_\snn, \obs-\bm \Phi_0^T\Kmat \hat \cvec-\epsilon\bm 1_\snn)\\
\hat {\bm \zeta} &=\max(\bm 0_\snn,-\obs+\bm \Phi_0^T\Kmat \hat \cvec-\epsilon\bm 1_\snn).
\end{align}
\end{subequations}
\change{Algorithm~\ref{algo:nonpar} lists the steps involved in the
  proposed nonparametric estimator.  } Note that the primal
\eqref{eq9} entails $(M+2)N$ variables whereas the dual
\eqref{eq:probnpd} has just $2N$. The latter can
be solved using sequential minimal optimization
algorithms~\cite{platt1999smo}, which here can afford simplified
implementation along the lines of e.g.,~\cite{kecman2003adatron}
because there is no bias term. However, for moderate problem
sizes ($<5,000$), interior point solvers are more
reliable~\cite[Ch.~10]{scholkopf2001} while having \emph{worst-case}
complexity ${\cal O} (N^{3.5})$. As a desirable byproduct, interior
point methods directly provide the Lagrange multipliers, which are
useful for recovering the primal variables (cf. \eqref{eq:dvec}).

\begin{algorithm}[t]                
  \caption{\change{Nonparametric batch PSD map estimator}}
  \label{algo:nonpar}    
  \begin{minipage}{20cm}
    \begin{algorithmic}[1]
      \STATE \textbf{Input:} $\{(\bm x_{n},
      \powsnn{\snind_n},\qensnn{\snind_n})\}_{n=1}^{N}$, $\{\txsgpsdsrn{\srind}(f)\}_{\srind=1}^M$ ,      $\epsilon$
      \STATE \textbf{Parameters:}
      $\lambda$, $\bm K(\bm x,\bm x')$
      \STATE $\bm \Phi = [\bm \phi (\bm x_1),\ldots, \bm \phi (\bm x_N)]$
      \STATE $\bm \Phi_0 = \bm I_N \odot \bm \Phi$

      \STATE Form $\bm K$, whose
      $(n,n')$-th  block is $\bm K (\bm x_n,\bm x_{n'})$
      \STATE $\bm K_0 =\bm \Phi_0^T \bm K \bm
      \Phi_0$ \STATE $\bm y =[y(\bm x_1),\ldots, y(\bm
        x_N)]^T$ \STATE Obtain $(\hat{\bm
        \alpha},\hat{\bm \beta})$ from
      \eqref{eq:probnpd}
      \STATE $[\hbm c_1^T ,\ldots, \hbm c_N^T]^T= [1/({2\lambda\snn})] \bm \Phi_0(\hat \alphavec- \hat {\bm  \beta} )$
      \STATE \textbf{Output:} Function 
      $\hat \Gamma({\snind},{f})
      =\sum_{\srind=1}^{\srn}\hat l_{\srind}({\snind})
      \txsgpsdsrn{\srind}(f)
      $, where\\ $ [\hat l_1(\bm x),\ldots,\hat l_M(\bm x)]^T=\sum_{n=1}^N \bm K (\bm x,\bm x_{n}) \hbm c_{n}$. 
    \end{algorithmic}
 	\end{minipage}
\end{algorithm}

\noindent
\subsection{Semiparametric regression using quantized data}
\label{sec:spr}

The nonparametric estimators in Secs.~\ref{sec:nprnq} and
\ref{sec:npr} are universal in the sense that they can approximate
wide classes of functions, including all continuous functions $\bm l$
vanishing at infinity, with arbitrary accuracy provided that the
number of measurements is large
enough~\cite{carmeli2010vector}. However, since measurements are
limited in number and can furthermore be quantized, incorporating
available prior knowledge is crucial to improve the accuracy of the
estimates. One could therefore consider applying \emph{parametric}
approaches since they can readily incorporate various forms of prior
information. However, these approaches lack the flexibility of
nonparametric techniques since they can only estimate functions in
very limited classes. Semiparametric alternatives offer a ``sweet
spot'' by combining the merits of both
approaches~\cite{scholkopf2001}.

This section presents semiparametric estimators capable of capturing
prior information about the propagation environment yet preserving the
flexibility of the nonparametric estimators in Secs.~\ref{sec:nprnq}
and \ref{sec:npr}. To this end, an estimate of the form $\ch\pd =
\ch'\pd + \check \ch\pd$ is pursued, where (cf. Secs.~\ref{sec:nprnq}
and~\ref{sec:npr}) the nonparametric component $\ch'\pd$ belongs to an
RKHS $\sspacep$ with kernel matrix $\bm K'$
; whereas the
parametric component is given by
\begin{equation}
\label{eq:wbemspan}
\check \ch (\bm x) = \sum_{\nu=1}^{N_B}
{\bm B}_\nu (\bm x) {\bm \theta}_\nu 
:= \bm B (\bm x) \bm \theta
\end{equation}
with $\bm B (\bm x) := [\bm B_1 (\bm x), \ldots, \bm B_{N_B} (\bm x)]$
collecting $N_B$ user-selected basis matrix functions $\bm B_\nu (\bm
x):\mathcal{R}\rightarrow \rfield^{M\times M}$, $ {\nu=1,\ldots,N_B}$,
and $\bm \theta:=[\bm \theta_1^T,\ldots, \bm \theta_{N_B}^T]^T$. 

If the transmitter locations $\{\bm \chi_m\}_{m=1}^M$ are
approximately known, the free space propagation loss can be described
by matrix basis functions of the form $\bm B_m (\bm x) = f_m (|| \bm x
- \bm \chi_m ||) \bm e_{M,m} \bm e_{M,m}^T$, where $f_m(|| \bm x - \bm
\chi_m ||)$ is the attenuation between a transmitter located at $\bm
\chi_m$ and a receiver located at an arbitrary point $\bm x$; see
Sec.~\ref{sec:sim} for an example. Note that if this basis accurately
captures the propagation effects in $\mathcal{R}$, then the $m$-th
entry of the estimated $\bm \theta_m$ is approximately proportional to
the transmit power of the $m$-th transmitter.

An immediate two-step approach to estimating $\bm l$ is to first fit
the data with $\check{\bm l}$ in \eqref{eq:wbemspan}, and then fit the
residuals with $\bm l'$ as detailed in Sec.~\ref{sec:npr}. Since this
so-termed \emph{back-fitting} approach is known to yield sub-optimal
estimates~\cite{scholkopf2001}, this paper pursues a joint fit, which
constitutes a novel approach in kernel-based learning for
vector-valued functions. To this end, define $\sspace$ as the space of
functions $\bm l$ (not necessarily an RKHS) that can be written as $
\bm l= \bm l' + \check{\bm l}$, with $\bm l' \in \cal H'$ and
$\check{\bm l}$ as in \eqref{eq:wbemspan}. One can thereby seek
semiparametric estimates of the form
\begin{equation}
\hat \ch = \argmin_{\ch \in \cal H} \frac{1}{N} \sum_{n=1}^N {\cal L}( y (\snind_n)- {\bm \phi}^T (\snind_n) \chsnn{\snind_n}) +\lambda \| \ch' \|_{\cal H'}^2\label{eq:rprobsp}
\end{equation}
where the regularizer involves only the nonparametric component
through the norm $||\cdot||_\sspacep$ in $\sspacep$. Using
\cite[Th. 3.1]{argyriou2014unifying}, one can readily generalize the
representer theorem in~\cite[Th. 4.3]{scholkopf2001} to the present
semiparametric case. This yields
 $\hat{\ch} (\bm x) =  
\bm K' (\bm x) \hat{\bm c}' + \bm B (\bm x) \hat{\bm \theta}$, where 
\begin{align}\label{eq16}
(\hat{\bm c}',\hat{\bm \theta}) =  \argmin_{\bm c',\bm \theta} \sum_{n=1}^N & {\cal L}\Big( y (\bm x_n)- {\bm \phi}^T (\bm x_n) [ \bm K' (\bm x_n)\bm c' \nonumber \\
&+ \bm B (\bm x_n) \bm \theta ]\Big) +\lambda N \bm c'^T \bm K' \bm c'\;.
\end{align}
Comparing \eqref{eq6} with \eqref{eq16}, and replacing $\bm K (\bm
x_n) \bm c$ 
 with $ \bm K' (\bm x_n)\bm c'+ \bm B (\bm x_n) \bm \theta$ yields
 the next result (cf. Proposition~\ref{prop:repth}).
\begin{proposition} \label{prop2}
The problem in \eqref{eq16} with $\mathcal{L}$ the
$\epsilon$-insensitive loss function ${\cal L}_{1\epsilon}$ can be
expressed as 
\begin{align}
(\hat{\bm c}',&\hat{\bm \theta},\hat{\bm \xi},\hat{\bm \zeta)} =
  \argmin_{\bm c',\bm \theta, \bm \xi, \bm \zeta} \sum_{n=1}^N
  (\xi_n +\zeta_n)+\lambda N \bm c'^T \bm K' \bm c' \nonumber \\ {\rm
    s. to}\;& \xi_n \geq y (\snind_n)- {\bm \phi}^T (\bm x_n) [\bm K'
    (\bm x_n)\bm c' + \bm B(\bm x_n) \bm \theta] - \epsilon,\:\xi_n
  \geq 0 \nonumber\\ &\zeta_n \geq - y (\snind_n) + {\bm \phi}^T (\bm
  x_n) [\bm K' (\bm x_n)\bm c' + \bm B(\bm x_n) \bm \theta] -
  \epsilon,\zeta_n \geq 0 \nonumber\\ &~~~n=1,\ldots,N. \label{eq17}
\end{align}
\end{proposition}

The primal problem in \eqref{eq17} entails vectors of size $MN$, which
motivates solving its dual version. Upon defining $\bm B$ as an $NM
\times N_BM$ matrix whose $(n,\nu)$-th block is $\bm B_{\nu} (\bm
x_n)$, and representing the Langrange multipliers associated with the
$\{\xi_n\}$ and $\{\zeta_n\}$ constraints by $\bm \alpha$ and $\bm
\beta$, the dual of \eqref{eq17} is
\begin{align}
	\label{eq:probnpdsp}
	\begin{aligned}
		&(\hat{\bm \alpha},\hat{\bm \beta}) = \argmin_{\bm \alpha,\bm \beta}
		&&\frac{1}{4\snn\lambda}(\alphavec-\bm \beta)^T \bm K_0' (\alphavec-\bm \beta )\\
		&&&-(\obs-\epsilon\bm 1_\snn)^T\alphavec +(\obs+\epsilon \bm 1_\snn)^T\bm \beta\\
		&\hspace{1.5cm}\st&& \bm 0_\snn\leq \alphavec\leq \bm 1_\snn,
		\bm 0_\snn\leq \bm \beta \leq \bm 1_\snn \\
		& {}&& \bm B^T\bm \Phi_0 (\bm \alpha - \bm \beta) = \bm 0
	\end{aligned}
\end{align}
where $\bm K'_0 := \bm \Phi_0^T \bm K' \bm \Phi_0$. Except for the
last constraint and the usage of $\bm K'_0$, \eqref{eq:probnpdsp} is
identical to~\eqref{eq:probnpd}. Similar to \eqref{eq:cvecopt}, the
primal vector of the nonparametric component can be readily obtained
from the KKT conditions as
\begin{equation}
	\label{eq21}
	\hat{\bm c}' = \frac{1}{2\lambda\snn} \bm \Phi_0(\hat \alphavec- \hat {\bm  \beta} ) \;.
\end{equation}
Although $\dvec$ can also be obtained from the KKT conditions, this
approach is numerically unstable. It is preferable to obtain $\dvec$
from the Lagrange multipliers of \eqref{eq:probnpdsp}, which are known
e.g. if an interior-point solver is employed. Specifically, noting
that \eqref{eq17} is the dual of \eqref{eq:probnpdsp}, it can be seen
that $\hat{\bm \theta}$ equals the multipliers of the last constraint
in \eqref{eq:probnpdsp}. \change{Algorithm~\ref{algo:semipar}
  summarizes the proposed semiparametric estimator.  }

\begin{algorithm}[t]                
  \caption{\change{Semiparametric batch PSD map estimator}}
  \label{algo:semipar}    
  \begin{minipage}{20cm}
    \begin{algorithmic}[1]
      \STATE \textbf{Input:} $\{(\bm x_{n},
      \powsnn{\snind_n},\qensnn{\snind_n})\}_{n=1}^{N}$,
      $\{\txsgpsdsrn{\srind}(f)\}_{\srind=1}^M$,       $\epsilon$
      \STATE \textbf{Parameters:}
$\lambda$, $\bm K'(\bm x,\bm x')$,  $\{\bm B_\nu
(\bm x)\}_{\nu=1}^{N_B}$
      \STATE $\bm \Phi = [\bm \phi (\bm x_1),\ldots, \bm \phi (\bm x_N)]$
      \STATE $\bm \Phi_0 = \bm I_N \odot \bm \Phi$

      \STATE Form $\bm K'$, whose
      $(n,n')$-th   block is $\bm K' (\bm x_n,\bm x_{n'})$
      \STATE $\bm K_0' =\bm \Phi_0^T \bm K' \bm
      \Phi_0$ \STATE $\bm y =[y(\bm x_1),\ldots, y(\bm
        x_N)]^T$ 
      \STATE Form $\bm B$, whose $(n,\nu)$-th block is $\bm B_{\nu} (\bm x_n)$
\STATE Obtain $(\hat{\bm
        \alpha},\hat{\bm \beta})$ from
      \eqref{eq:probnpdsp} and set $\hbm \theta:=[{\hbm{\theta}}_1^T,\ldots,
        \hbm \theta_{N_B}^T]^T$  to\\be the optimal
      Lagrange multiplier of the  last constraint
      \STATE $[\hbm c_1^T ,\ldots, \hbm c_N^T]^T= [1/({2\lambda\snn})] \bm \Phi_0(\hat \alphavec- \hat {\bm  \beta} )$
      \STATE \textbf{Output:} Function 
      $\hat \Gamma({\snind},{f})
      =\sum_{\srind=1}^{\srn}\hat l_{\srind}({\snind})
      \txsgpsdsrn{\srind}(f)
      $, where\\ \hspace{-.8cm}$ [\hat l_1(\bm x),\ldots,\hat l_M(\bm x)]^T=
\sum_{n=1}^N \bm K' (\bm x,\bm x_{n}) \hbm c_{n}+
\sum_{\nu=1}^{N_B}
{\bm B}_\nu (\bm x) {\hbm \theta}_\nu 
$. 
    \end{algorithmic}
 	\end{minipage}
\end{algorithm}

\subsection{Regression with conditionally positive definite kernels}

So far, the kernels were required to be positive definite.  This
section extends the semiparametric estimator in Sec.~\ref{sec:spr} to
accommodate the wider class of \emph{conditionally positive definite}
(CPD) kernels.  CPD kernels are natural for estimation problems that
are invariant to translations in the data~\cite[p. 52]{scholkopf2001},
as occurs in spectrum cartography.
Accommodating CPD kernels also offers a generalization of thin-plate
splines (TPS), which have well-documented merits in capturing
shadowing of propagation channels~\cite{bazerque2011splines}, to
operate on quantized data.

Consider the following definition, which generalizes that of scalar
CPD kernels~\cite[Sec.~2.4]{scholkopf2001}. Recall that, given $\{ \bm
x_n\}_{n=1}^N$, $\bm K$ is a matrix whose $(n,n')$-th block is $\bm K
(\bm x_n,\bm x_{n'})$.
\begin{definition}
A kernel
$\kermat(\snind_1,\snind_2):\rfield^\dimn\times\rfield^\dimn\rightarrow\rfield^{\srn\times\srn}$
is CPD with respect to $\{\bm B_\nu (\bm x)\}_{\nu=1}^{N_B}$ if it
satisfies $\cvec^T \Kmat \cvec\geq 0$ for every finite set $\{ \bm
x_n\}_{n=1}^N$ and all $\cvec$ such that $\Psimat^T\cvec=\bm 0$.
\end{definition}
Observe that any positive definite kernel is also CPD since it
satisfies $\cvec^T \Kmat \cvec\geq 0$ for all $\cvec$.

To see how CPD kernels can be applied in semiparametric regression, note
that \eqref{eq21} together with the last constraint in
\eqref{eq:probnpdsp} imply that the solution to \eqref{eq17} satisfies
$\bm B^T \hat{\bm c}' = \bm 0$. Therefore, \eqref{eq17} can be
equivalently solved by confining the vectors $\bm c'$ to lie in the
null space of $\bm B^T$:
\begin{align}
(\hat{\bm c}',&\hat{\bm \theta},\hat{\bm \xi},\hat{\bm \zeta)} =  \argmin_{\bm c',\bm \theta, \bm \xi,\bm \zeta} \bm 1_N^T (\bm \xi+\bm \zeta)+\lambda N \bm c'^T \bm K' \bm c' \nonumber  \\
{\rm s.t.}\;& \bm \xi  \geq \bm y - {\bm \Phi_0}^T (\bm K' \bm c' + \bm B \bm \theta) - \epsilon \bm 1_N\:,\;\;\;\;\bm \xi \geq \bm 0_N \nonumber\\
&\bm \zeta \geq - \bm y + {\bm \Phi_0}^T 
(\bm K' \bm c' + \bm B \bm \theta) - \epsilon \bm 1_N\:,\;\bm \zeta \geq \bm 0_N \nonumber\\
&\bm B^T \bm c' = \bm 0. \label{eq22}
\end{align}
The new equality constraint ensures that the objective of \eqref{eq22}
is convex in the feasible set if $\bm K'(\bm x,\bm x')$ is CPD with
respect to $\{\bm B_\nu (\bm x)\}_{\nu=1}^{N_B}$. However,
\eqref{eq22} is susceptible to numerical issues because the block
matrix $\bm K'$ may not be positive semidefinite. One can circumvent
this difficulty by adopting a change of variables $\bm c' := \proj
\tcvec$, where $\proj := \bm I_{\srn\snn} -
\Psimat(\Psimat^T\Psimat)\inv \Psimat^T \in
\rfield^{\srn\snn\times\srn\snn}$ is the orthogonal projector onto the
null space of $\Psimat^T$. By doing so,  \eqref{eq22} becomes
\begin{align}
(\hat{\tcvec},&\hat{\bm \theta},\hat{\bm \xi},\hat{\bm \zeta)} =  \argmin_{\tcvec,\bm \theta, \bm \xi,\bm \zeta} \bm 1_N^T (\bm \xi+\bm \zeta)+\lambda N \tcvec^T \proj \bm K' \proj \tcvec \nonumber  \\
{\rm s.t.}\;& \bm \xi  \geq \bm y - {\bm \Phi_0}^T (\bm K' \proj \tcvec + \bm B \bm \theta) - \epsilon \bm 1_N,\;\;\;\;\bm \xi \geq \bm 0_N \nonumber\\
&\bm \zeta \geq - \bm y + {\bm \Phi_0}^T 
(\bm K' \proj \tcvec + \bm B \bm \theta) - \epsilon \bm 1_N\:,\;\bm \zeta \geq \bm 0_N \label{eq22b}
\end{align}
where $\proj\bm K' \proj$ is guaranteed to be positive semidefinite.

A similar argument applies to the dual formulation in
\eqref{eq:probnpdsp}, where, for feasible $\bm \alpha$ and $\bm \beta$,
it holds that $(\bm \alpha - \bm \beta)^T \bm K_0'(\bm \alpha - \bm
\beta)\geq 0$ with $\bm K'(\bm x,\bm x')$ CPD. To avoid numerical
issues, observe that any feasible $\bm \alpha ,\bm \beta$ satisfy $\bm
\Phi_0 (\bm \alpha - \bm \beta ) = \proj \bm \Phi_0 (\bm \alpha - \bm
\beta )$, and thus \eqref{eq:probnpdsp} can be equivalently expressed as
\begin{align}
\label{eq23}
\begin{aligned}
&(\hat{\bm \alpha},\hat{\bm \beta}) = \argmin_{\bm \alpha,\bm \beta}
&&\frac{1}{4\snn\lambda}(\alphavec-\bm \beta)^T \tilde{\bm K} (\alphavec-\bm \beta )\\
&&&-(\obs-\epsilon\bm 1_\snn)^T\alphavec +(\obs+\epsilon \bm 1_\snn)^T\bm \beta\\
&\hspace{1.5cm}\st&& \bm B^T\bm \Phi_0 (\bm \alpha - \bm \beta) = \bm 0\\
& {}&& \bm 0_\snn\leq \alphavec\leq \bm 1_\snn,
\bm 0_\snn\leq \bm \beta \leq \bm 1_\snn 
\end{aligned}
\end{align}
where $\bm K_0'$ has been replaced with the positive semidefinite
matrix $\tilde{\bm K} := \bm \Phi_0^T \proj \bm K' \proj \bm \Phi_0$.
With this reformulation, although the optimal $\hat{\bm c}'$ can still
be recovered from \eqref{eq21}, $\hat{\dvec}$ can no longer be
obtained as the Lagrange multiplier $\bm \mu$ of the equality
constraint in \eqref{eq23}. This is due to the change of variables,
which alters \eqref{eq23} from being the dual of \eqref{eq22b}. As
derived in Appendix~\ref{app:dvec}, $\hat{\dvec}$ can instead be
recovered as
\begin{align}
\hat{\dvec} = \bm \mu -  (\Psimat^T\Psimat) \inv \Psimat^T \Kmat'  \hat{ \cvec}' \label{eq:dvec}.
\end{align}

Broadening the scope of semiparametric regression to include CPD
kernels leads also to generalizations of TPS -- arguably the most
popular semiparametric interpolator, which derives its name because
the TPS estimate mimics the shape of a thin metal plate that minimizes
the bending energy when anchored to the data
points~\cite{bookstein1989principalwarps}. With $\bm x
\define[x_1,\ldots,x_d]^T$, TPS adopts the parametric basis $
\wbasiswbasisnsnn{1}{\snind} = \bm I_\srn,~
\wbasiswbasisnsnn{2}{\snind} = x_1\bm I_\srn,\ldots,
\wbasiswbasisnsnn{1+\dimn}{\snind} = x_\dimn\bm I_\srn$, and the diagonal
matrix kernel
\begin{align}
	\label{eq:kermat}
	\kermat'(\snind_1,\snind_2) = r( ||\snind_1-\snind_2||_2^2  )\bm I_\srn
\end{align}
where $r (z)$ denotes the radial basis function 
\begin{align}
	r(z) := \begin{cases}
		z^{2s-\dimn}\log(z)&\text{if}~\dimn~\text{is even}\\
		z^{2s-\dimn}&\text{otherwise}
	\end{cases}
\end{align}
for a positive integer $s$ typically set to
$s=2$~\cite[eq. (2.4.9)]{wahba1990splinemodels},\cite{bazerque2011splines}.
The kernel in \eqref{eq:kermat} can be shown to be CPD with respect to
$\{\bm B_\nu (\bm
x)\}_{\nu=1}^{1+d}$~\cite[p. 32]{wahba1990splinemodels}.  The norm in
the RKHS $\sspacep$ induced by \eqref{eq:kermat}, which can be
evaluated as in Sec.~\ref{sec:nprnq}, admits the equivalent form
\[
||\ch'||_\sspacep^2 =
	\sum_{\srind=1}^{\srn}\int_{\rfield^\dimn}||\nabla^2 \weln{\srind}(\bm
	z)||_F^2d\bm z
\]
where $||\cdot||_F$ denotes Frobenius norm, and $\nabla^2$ the
Hessian. Therefore, the RKHS norm of $\sspacep$ captures the
conventional notion of smoothness embedded in the magnitude of the
second-order derivatives.  Among other reasons, TPS are popular
because they do not require parameter tuning, unlike e.g., Gaussian
kernels, which need adjustment of their variance parameter. The
novelty here is the generalization of TPS to vector-valued function
estimation from quantized observations. {Unlike
  \cite{bazerque2011splines}, which relies on un-quantized
  periodograms, the proposed scheme is based on quantized power
  measurements. }

\subsection{Multiple measurements per sensor}
\label{sec:emms}
For simplicity, it was assumed so far that each sensor collects and
reports a single measurement to the fusion center. However, $P > 1$
measurements can be obtained per sensor by changing the filter impulse
response between measurements, or, by appropriately modifying the
compression matrix in their A2ICs; cf. Remark~\ref{remark:a2ic}. A
naive approach would be to regard the $P$ measurements per sensor as
measurements from $P$ different sensors at the same location. However,
this increases the problem size by a factor of $P$, and one has to
deal with a rank deficient kernel matrix $\bm K$ of dimension $MNP$,
which renders the solutions of~\eqref{eq6p},~\eqref{eq6} and
\eqref{eq16} non-unique.

A more efficient means of accommodating $P$ measurements per sensor
in the un-quantized scenario is to reformulate \eqref{eq4}~as
\begin{equation}
\hat \ch = \argmin_{\ch \in \cal H} \frac{1}{NP}\sum_{n=1}^N
\sum_{p=1}^P {\cal L}( \hat \pi_p (\snind_n)- {\bm \phi}_p^T (\snind_n) \chsnn{\snind_n}) +\lambda \| \ch \|_{\cal H}^2 \label{eq26}
\end{equation}
where $\hat \pi_{\mpsnind}({\snind_n})$ and
$\msnnmpsnn{\snind_n}{\mpsnind}$ correspond to the $\mpsnind$-th
measurement reported by the sensor at location $\snind_n$. Collect all
observations in $\hat{\bbm{\pi}} :=
[\hat{\pi}_{1}({\snind_1}),\hat{\pi}_2({\snind_1}),\ldots,\hat{\pi}_{\mpsnn}(
  {\snind_\snn})]^T$, let $\bar{\bm \Phi} :=
[\msnnmpsnn{\snind_1}{1},\msnnmpsnn{\snind_1}{2},\ldots,\msnnmpsnn{\snind_\snn}{\mpsnn}]$,
and let $\bar{\bm \Phi}_0 := (\bm I_N \otimes {\bm 1}_P^T) \odot \bar{\bm
  \Phi} \in \mathbb{R}^{MN \times NP}$.  {As before, the
  representer theorem implies that the minimizer of \eqref{eq26} is
  given by $\hbm l(\bm x) = \sum_{n=1}^N \bm K (\bm x,\bm x_n)
  \hat{\bbm c}_n:= \bm K(\bm x) \hb{\bm c}$. If $\mathcal{L}$ is the
  square loss $\mathcal{L}_2$ (see Sec.~\ref{sec:nprnq}), then
\begin{equation}
\label{eq:optcridgeregressionmultiplemeasurements}
\hb{\bm c}  = (\bbm \Phi_0 \bbm \Phi_0^T \bm K+\lambda N\bm I_{MN})\inv
\bbm \Phi_0 \hb{\bm \pi}
\end{equation}
which generalizes \eqref{eq:optcridgeregression} to $P\geq 1$.  } Note
however that $\bar{ \bm c}$ has dimension $MN$, whereas the
aforementioned naive approach would result in $MNP$. Likewise, $\bm K$
is of size $MN \times MN$.

If $ \hat \pi_p (\snind_n)$ is replaced with
$\obssnnmpsnn{\snind_n}{p}$ in \eqref{eq26}, the resulting expression
extends \eqref{eq4b} to multiple measurements per sensor.  If
$\mathcal{L}$ is the $\epsilon$-insensitive cost
$\mathcal{L}_{1\epsilon}$, such an expression is minimized for $\hbm
l(\bm x) = \bm K(\bm x) \hb{\bm c}$ with
\begin{align}
(\hat{\bar{\bm c}},\hat{\bar{\bm \xi}}, \hat{\bar{\bm \zeta}}) & =  
\argmin_{\bar{\bm c},\bar{\bm \xi},\bar{\bm \zeta}} \;
\bm 1_{NP}^T (\bar{\bm \xi} + \bar{\bm \zeta})+\lambda NP \bar{\bm c}^T \bm K \bar{\bm c} \nonumber  \\
{\rm s.~to}\;\;& \bar{\bm \xi}  \geq \bar{\bm y} - \bar{\bm \Phi}_0^T \bm K \bar{\bm c} - \epsilon \bm 1_{NP},\;\;\;\;\bar{\bm \xi} \geq \bm 0_{NP} \nonumber\\
& \bar{\bm \zeta} \geq - \bar{\bm y} + \bar{\bm \Phi}_0^T 
\bm K \bar{\bm c} - \epsilon \bm 1_{NP}\:,\:\bar{\bm \zeta} \geq \bm 0_{NP} \label{eq27}
\end{align}
where $\bar \obs :=
[\obssnnmpsnn{\snind_1}{1},\obssnnmpsnn{\snind_1}{2},\ldots,\obssnnmpsnn{\snind_\snn}{\mpsnn}]^T$.
Note that \eqref{eq27} reduces to~\eqref{eq9} if $P=1$.  The dual
formulation is the same as \eqref{eq:probnpd}, except that $\bm
\alpha$, $\bm \beta$, $\bm \Phi_0$, and $N$ are replaced with $\bar{\bm
  \alpha}$, $\bar{\bm \beta}$, $\bar{\bm \Phi}_0$, and $NP$,
respectively. The primal solution can be recovered as $\hat{\bar{\bm
    c}} = (2\lambda NP)^{-1} \bar{\bm \Phi}_0(\hat{\bar{\bm \alpha}} -
\hat{\bar {\bm \beta}})$; cf.~\eqref{eq:cvecopt}.

The multi-measurement version of the semiparametric estimator in
\eqref{eq:rprobsp} is
\begin{equation}
\hat \ch = \argmin_{\ch \in \cal H} \frac{1}{NP}\sum_{n=1}^N 
\sum_{p=1}^P {\cal L}( y_p (\snind_n)- \bm \phi_p^T (\snind_n) \chsnn{\snind_n}) +\lambda \| \ch' \|_{\cal H'}^2  \label{eq28}
\end{equation}
and the counterpart to~\eqref{eq17} is obtained by replacing
$\bm \xi$, $\bm \zeta$, $\bm y$, $\bm \phi (\bm x_n)$ and $N$, with $\bar
{\bm \xi}$, $\bar{\bm \zeta}$, $\bar{\bm y}$, $\bm \phi_p (\bm x_n)$ and $NP$, respectively. Likewise, the dual formulation is
obtained by substituting $\bm \alpha$, $\bm \beta$, $\bm
  \Phi_0$, $N$, and $\tbm K$ in~\eqref{eq23} with $\bar{\bm
  \alpha}$, $\bar{\bm \beta}$, $\bar {\bm \Phi}_0$, $NP$,
and
\begin{align}
\label{eq:tbbRdef}
\tilde {\bar {\bm K}} := \bar{\bm \Phi}_0^T \proj \bm K'
\proj \bar{\bm \Phi}_0
\end{align}
respectively. The primal variables are recovered again as 
$\hat{\bar{\bm c}} = (2\lambda NP)^{-1} \bar{\bm \Phi}_0 (\hat{ \bar{\bm \alpha}} - \hat{\bar{\bm \beta}})$ and $\hat{\dvec} = \bar {\bm \mu} - (\Psimat^T\Psimat) \inv \Psimat^T \Kmat' \hat{\bar{\cvec}}$, where 
$\bar{\bm \mu}$ is the Lagrange multiplier vector associated with the equality constraints in the dual problem; cf.~\eqref{eq:dvec}.

\noindent
\begin{remark}[Non-uniform quantization] 
\label{remark:nonuniform}
Unless $\hat \pi({\snind_n})$
is uniformly distributed, non-uniform quantization may be preferable
over the uniform quantization adopted so far. With $R$ quantization
regions specified by the boundaries
$0=\qreglim_0<\qreglim_1<\ldots<\qreglim_\qregn$, the quantized
measurements are $ \pi_Q({\snind_n}):=Q(\hat \pi({\snind_n}))=i$ for
$\hat \pi({\snind_n}) \in [\tau_{i},\tau_{i+1})$. The general
  formulations~\eqref{eq26} and~\eqref{eq28} can accommodate
  non-uniformly quantized observations by replacing $\epsilon$ in all
  relevant optimization problems with 
  $\epsilon(\bm x_n) := (\tau_{\pi_Q({\snind_n})+1}
  -\tau_{\pi_Q({\snind_n})})/2$; and likewise modifying the centroid
  expression from $y(\bm x_n) := [2\pi_Q({\snind_n})+1]\epsilon$ to $y
  (\bm x_n) :=
  (\tau_{\pi_Q({\snind_n})+1}+\tau_{\pi_Q({\snind_n})})/2$.
\end{remark}

\noindent
\begin{remark}[Enforcing nonnegativity]  Since all $M$ entries of
vector $\bm l\pd$ represent power, they are inherently nonnegative. To
exploit this information, at least partially, one can enforce
non-negativity of $\bm l\pd$ at all sensor locations by introducing
the constraint $\bm K (\bm x_n) \bm c \ge {\bm 0}_{M}$ for
$n=1,\ldots,N$ in~\eqref{eq9}. Another approach is to include $M$
``{virtual measurements}'' for every sensor location $\snind_n$ to
promote estimates $\ch(\bm x)$ satisfying $0\leq l_m(\bm
x_n)<\qreglim_R$ for all $m$. In this way, the estimation algorithm
no longer uses the set of ``real'' measurements $\{\big( y_{p}(\bm
x_{n}), \bm \phi_{p}(\snind_n),\epsilon_{p}(\bm x_n)\big)\}_{p=1}^P$
for every sensor $n=1,\ldots, N$, where $\epsilon_{p}(\bm x_n)$ is the
quantization interval of the $p$-th measurement obtained by the $n$-th
sensor. Instead, it uses its extended version $\{\big( y_{p}(\bm
x_{n}), \bm \phi_{p}(\snind_n),\epsilon_{p}(\bm
x_n)\big)\}_{p=1}^{P+M}$, where $\bm \phi_{P+m}(\snind_n)=\bm
e_{M,m}$, $y_{P+m}(\bm x_{n}):=(\tau_0 +\tau_R)/2$,
$\epsilon_{P+m}(\bm x_n) :=( \tau_R -\tau_0)/2$ for $m=1,\ldots,M$.
The measurements $p=P+1,\ldots,P+M$ are termed ``virtual'' since they
are appended to the ``real'' measurements by the fusion center, but
are not acquired by the sensors. Note that this approach does not
constrain $\ch$ to be entry-wise non-negative at the sensor locations,
it just promotes estimates satisfying this condition.
\end{remark}

{
\begin{remark}[Computational complexity]
\label{remark:computationalcomplexity}
With un-quantized data, the estimate
\eqref{eq:optcridgeregressionmultiplemeasurements} requires
$\mathcal{O}(M^3N^3+PM^2N)$ operations.  With nonparametric estimation
from quantized data, solving the dual of \eqref{eq27} through interior
point methods takes $\mathcal{O}((NP)^{3.5})$ iterations. A similar
level of complexity is incurred by its semiparametric
counterpart. Albeit polynomial, this complexity may be prohibitive in
real-time applications with limited computational capabilities if the
number of measurements $NP$ is large.  For such scenarios, an online
algorithm with linear complexity is proposed in Sec.~\ref{sec:oi},
which is guaranteed to converge to the nonparametric estimate
\eqref{eq6}.  An online algorithm for semiparametric estimation can be
found in \cite{romero2015onlinesemiparametric}.
\end{remark}
}

\change{
\begin{remark}
\label{remark:timescale}
In real applications, low SNR conditions, transmission beamforming,
and the hidden terminal problem may limit the quality of PSD map
estimates relying on short observation windows. Thus, highly accurate
estimates may require longer observation intervals to average out the
undesirable effects of noise and small-scale fading. This constitutes
a tradeoff between estimation accuracy and temporal resolution of PSD
maps that is inherent to the spectrum cartography problem. This
tradeoff may not pose a difficulty in applications such as in TV
networks, where transmitters remain active or inactive for long time
intervals, typically months or years, and therefore a high temporal
resolution is unnecessary. In other scenarios, it may suffice to know
whether the primary users are active or inactive, in which case a high
estimation accuracy is not needed, and therefore short observation
windows may be enough. However, in those scenarios where both high
accuracy and fine temporal resolution are required, one would need to
deploy more sensors.  The present paper alleviates the negative impact
of the aforementioned tradeoff by reducing the required communication
bandwidth. This brings a threefold benefit: (i) in a fixed time
interval, each sensor may report more measurements to the fusion
center, thus improving averaging and therefore the accuracy of the
estimates; (ii), a larger number of sensors can be deployed; and
(iii), the latency of the communication between sensors and fusion
center is reduced.
\end{remark}
}

\section{Online Algorithm}
\label{sec:oi}

The algorithms proposed so far operate in batch mode, thus requiring
all observations before they can commence. Moreover, their
computational complexity increases faster than linearly in $NP$, which
may be prohibitive if the number of measurements is large relative to
the available computational resources. These considerations motivate
the development of online algorithms, which can both approximate the
solution of the batch problem with complexity $\mathcal{O}(NP)$ and
update $\hbm l\pd$ as new measurements arrive at the fusion center.
Online algorithms are further motivated by their ability to track a
time-varying $\ch$.

Although online algorithms can be easily constructed by iteratively
applying batch algorithms over sliding
windows~\cite{sebald2000equalization}, online strategies with
instantaneous updates are preferred~\cite{diethe2013review}.  An
elegant approach for kernel-based learning relying on stochastic
gradient descent (SGD) in the RKHS is developed
in~\cite{kivinen2004online} for \emph{scalar} kernel machines. Its
counterpart for vector-valued functions is described
in~\cite{audiffren2013operatorvalued}, but it is not directly
applicable to the present setup since it requires differentiable
objectives and vectors $\m(\snind_\nu)$ that do not depend on
$\nu$. This section extends the algorithm
in~\cite{audiffren2013operatorvalued} to accommodate the present
scenario.

For a selected ${\mathcal{L}}$, consider the instantaneous regularized
cost, defined for generic $\w\in \sspace$, $\snind_\nu$, $\m(\snind_\nu)$,
 and $\obsel(\snind_\nu)$ as
\begin{align}
\label{eq:instregerrdef}
\instregerr(\w,\m(\snind_\nu),\snind_\nu,\obsel(\snind_\nu)):=\mathcal{L}(\obsel(\snind_\nu)-\m^T(\snind_\nu)
\wsnn{\snind_\nu}) + \lambda \|\w\|_{\sspace}^2.
\end{align}
Note that \eqref{eq4} indeed minimizes the sample average of
$\instregerr$. Suppose that at time index $t=1,2,\ldots$ the fusion
center processes one measurement from sensor $n_t\in
\{1,\ldots,N\}$. If the fusion center uses multiple observations, say
$P$, from the $n'$-th sensor, then $n_{t_1}=n_{t_2}=\ldots=n_{t_P}=n'$,
where $\{t_1,\ldots,t_P\}$ depends on the fusion center schedule.

Upon processing the $t$-th measurement, the SGD update is
\begin{align}
\label{eq:upd}
\w\atstn{\stind+1}(\bm x)= \w\atstn{\stind}(\bm x) -\step \partial_{\w}\instregerr(\w\atstn{\stind},\msnn{\snindtt},\snindtt,\obssnn{\snindtt})(\bm x)
\end{align}
where $\w^{(t)}$ is the estimate at time $t$ before $y(\snind_{n_t})$
has been received; $\mu_t > 0$ is the learning rate; and
$\partial_{\w}$ denotes subgradient with respect to $\w$. In general,
$\step$ can be replaced with a matrix $\bm M_t$ to increase
flexibility. For $\mathcal{C}$ as in \eqref{eq:instregerrdef}, it can
be shown using~\cite[eq.~(2)]{audiffren2013operatorvalued} that
\begin{align}
\label{eq:gradrinst}
&\partial_{\w} \instregerr(\w,\m(\snind_\nu),\snind_\nu,\obsel(\snind_\nu))(\snind) 
\\ &=-\mathcal{L}'\big(\obsel(\snind_\nu)-\m^T(\snind_\nu)
\w(\snind_\nu)\big)
\bm K(\snind, \snind_\nu)\m(\snind_\nu)+2\lambda\w(\snind)
\nonumber
\end{align}
 and $\mathcal{L}'$ is a subgradient of $\mathcal{L}$, which for
 $\mathcal{L}_{1\epsilon}(e_\nu):=\max(0,|e_\nu|-\epsilon)$ is e.g.:
\begin{align}
\mathcal{L}_{1\epsilon}'(e_\nu) = \frac{ {\rm sgn}(e_\nu-\epsilon) + {\rm sgn}(e_\nu + \epsilon) }{2}.
\end{align}
Substituting \eqref{eq:gradrinst} into \eqref{eq:upd} yields
\begin{align}
\w\atstn{\stind+1}&(\snind)= (1-2\step\lambda)\w\atstn{\stind}(\snind)
\label{eq:wstnstind}\\&+\step\mathcal{L}'\big(\obssnn{\snindtt}-\msnnT{\snindtt}
\wsnnstn{\snindtt}{\stind}\big)\bm K(\snind, \snindtt)\msnn{\snindtt}.
\nonumber
\end{align}
Upon setting $\w\atstn{1}=\bm 0$, it follows that 
\begin{align}
\label{eq:oldec}
\w\atstn{\stind}(\snind) =
\sum_{i=1}^{\stind-1}\bm K(\snind,\snind_{n_i})\cvec\atstn{\stind}_i
\end{align}
for some $\cvec\atstn{\stind}_i$, $i=1,\ldots,t-1$. Interestingly,
although the representer
theorem~\cite[Thm.~5]{micchelli2005vectorvalued} has not been invoked,
the estimates $\w\atstn{\stind}(\snind)$ here have the form of those
in Sec.~\ref{sec:cl}.

If measurements come from sensors at different locations, the
functions $\bm K(\snind,{\snind_{n_i}})$, $i=1,\ldots,t$, are linearly
independent for properly selected kernels, and substituting
\eqref{eq:oldec} into \eqref{eq:wstnstind} results in the following
update rule:
\begin{align*}
  \cvec\atstn{\stind+1}_i &= \begin{cases}
    (1-2\step\lambda)\cvec_{i}\atstn{\stind}~\text{if}~i=1,\ldots,\stind-1\\
    \step\mathcal{L}'[\obssnn{\snindtt}-\msnnT{\snindtt} \wsnnstn{\snindtt}{\stind}]  \msnn{\snindtt}~\text{if}~i=t.
    \end{cases}
\end{align*}
 This equation reveals that the number of coefficients maintained
 increases linearly in $t$. This is the so-called \emph{curse of
   kernelization}~\cite{wang2012breaking}. However, if $\mu_t \lambda
 \in (0,1)$, then the amplitudes of the entries in $\bm c_{i}\atstn{\stind}$ are
 shrunk by a factor  $|1-2 \mu_t \lambda| < 1$. This justifies
 truncating \eqref{eq:oldec} as
\begin{align}
\w\atstn{\stind}(\snind) = \sum_{i=\max(1,\stind-\nterms)}^{\stind-1}\bm K(\snind,\snind_i)\cvec\atstn{\stind}_i
\end{align}
for some $\nterms>1$. On the other hand, if the fusion center processes multiple
observations per sensor, $\{\bm K(\snind,{\snind_{n_i}})\}_{i=1}^t$
are no longer linearly independent. In such a case, up to $N$ kernels
$\{\bm K(\snind,{\snind_{n}})\}_{n=1}^N$ are linearly independent,
which yields
\begin{align}
\label{eq:oldec22}
\w\atstn{\stind}(\snind) = \sum_{n = 1}^N\bm K(\snind, \snind_n)\cvec_{n}\atstn{\stind}.
\end{align}
After receiving the observation
$(\snindtt,\obssnn{\snindtt},\msnn{\snindtt})$ at time $t$, it follows
from \eqref{eq:oldec22} that one must obtain $\{
\cvec\atstn{\stind+1}_n\}_{n=1}^N$ as
\begin{align*}
  \cvec\atstn{\stind+1}_n &=
\begin{cases}
  (1-2\step\lambda)\cvec\atstn{\stind}_n&\text{if}~n\neq n_t\\
  (1-2\step\lambda)\cvec\atstn{\stind}_{n_t} + \step\mathcal{L}'\big[\obssnn{\snindtt}\\
    \hspace{.8cm}-\msnnT{\snindtt} \wsnnstn{\snindtt}{\stind}\big] \msnn{\snindtt}&\text{if}~n=n_t.
  \end{cases}
\end{align*}
Convergence of these recursions is characterized by the next
result, which adapts~\cite[Thm. 1]{audiffren2013operatorvalued} to the
proposed setup.

\newcommand{\co}{a_1}
\newcommand{\ct}{a_2}

\begin{theorem}
\label{thm:conv}
If $\lambda_{\max}(\bm K(\snind,\snind))<\bar{\lambda}^2<\infty$ for all
$\snind$, $ ||\m({\snindtt})||_2\leq \mbound$ for all $\stind$, and
$\step := \mu t^{-1/2}$ with $\mu \lambda < 1$, then the iterates in
\eqref{eq:wstnstind} with $\mathcal{L} = \mathcal{L}_{1\epsilon}$ satisfy
\begin{align}
\label{eq:regretbound}
&\frac{1}{\stn} \sum_{\stind=1}^\stn\instregerr(\w\atstn{\stind},\m({\snindtt}),\snindtt,\obsel({\snindtt}))\\
&\leq\inf_{\w\in \sspace}\left[
  \frac{1}{\stn}\sum_{\stind=1}^\stn\instregerr(\w,\m({\snindtt}),\snindtt,\obsel({\snindtt}))
  \right] + \frac{\co}{\sqrt{\stn}}+\frac{\ct}{{\stn}}  \nonumber
\end{align}
where $\ct \define \bar{\lambda}^2 \mbound^2/(8\lambda^2\mu)$ and
$\co\define 4(\bar{\lambda}^2\mbound^2\mu
+\ct)$.
\end{theorem}

\noindent {\bf Proof:} {See Appendix~\ref{sec:proof:thm:conv}. }\myQED

In words, Theorem \ref{thm:conv} establishes that the averaged
instantaneous error from the online algorithm converges to the
regularized empirical error of the batch solution.

\section{Numerical Tests}
\label{sec:sim}
\begin{figure*}[bhtp]
\centering
\includegraphics[height=8cm]{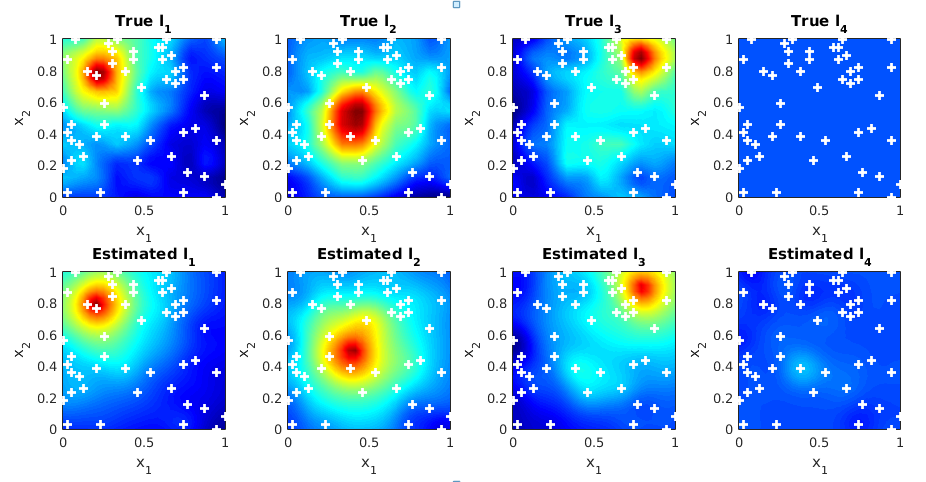}
\caption{{True and estimated functions $l_m(\bm x)$ with $\snn=50$,
    $\mpsnn=8$, $5$-bit quantization with CPQ, $\lambda = 10^{-9}$,
    and nonnegativity enforced through virtual sensors. Each column
    corresponds to the power of a channel. White crosses denote sensor
    locations. The PSD at any location can be reconstructed by
    substituting the values of these functions into
    \eqref{eq:rxsgpsdsnn}}.  }
\label{fig:maps}
\end{figure*}

In this section, the proposed algorithms are validated through numerical
experiments.  Following \cite{gudmundson1991correlation}, a correlated
shadow fading model was adopted, where, for $m=1,\ldots,M-1$,
\begin{align}
  10\log\chsrnsnn{\srind}{\snind} = 10\log{ A_\srind}  
-\gamma\log({\delta + \| \snind - \bm \chi_m \|})+\shadowing_\srind(\snind).
\end{align}
Here, $\delta>0$ is a small constant ensuring that the argument of the
logarithm does not vanish, $\gamma=3$ is the pathloss exponent, and
the parameters $A_\srind$ and $\bm \chi_m$ denote the power and
location of the $m$-th transmitter, respectively. The random shadowing
component $\shadowing_\srind(\snind)$ is generated as a zero-mean
Gaussian random variable with
$\expected{\shadowing_\srind(\snind)\shadowing_\srind(\snind')}=\sigma_s^2
\rho^{-||\snind-\snind'||}$, where $\sigma^2_s=2$ and $\rho=0.8$.  The
noise power was set to $\chsrnsnn{\srn}{\snind}=0.75$.

The $\snn$ sensors, deployed uniformly at random over the region
of interest, report $\mpsnn$ quantized measurements
$\{\qenestsnnmpsnn{\snind_n}{\mpsnind}
=\Q(|\powflsnnmpsnn{\snind_n}{\mpsnind}+\eta_p (\bm x_n)|)
=\Q(|\msnnmpsnnT{\snind_n}{\mpsnind} \chsnn{\snind_n}+\eta_p (\bm
x_n)|)\}_{p=1}^P$ to the fusion center, where $\eta_p(\bm x_n)
\sim\mathcal{N}(0,\sigma_\eta^2)$ simulates noise due to  finite
sample estimation effects.  The entries of
$\msnnmpsnn{\snind_n}{\mpsnind}$ were generated uniformly over the
interval $[0,1]$ for all $\snind_n$ and~$\mpsnind$. 

Two quantization schemes were implemented. Under uniform quantization (UQ),
the range of $\powflsnnmpsnn{\snind}{\mpsnind}$ was first determined
using Monte Carlo runs and the quantization region boundaries
$\qreglim_0<\qreglim_1<\ldots<\qreglim_\qregn$ were set such that $\qreglim_{i+1}-\qreglim_i =
2\epsilon\ \forall i$, where $\epsilon$ was such that the probability
of clipping ${\rm Pr}\{\powflsnnmpsnn{\snind}{\mpsnind}>\qreglim_\qregn\}$
was approximately $10^{-3}$ and $\qregn := 2^b$, with $b$ the number of
bits per measurement. Under constant probability
quantization~(CPQ), these boundaries were chosen such that
${\rm Pr}\{\qreglim_{i}\leq
  \powflsnnmpsnn{\snind}{\mpsnind}<\qreglim_{i+1}\}$ was approximately
constant for all $i$.

The true PSD map in the region $[0,1] \times [0,1] \subset \rfield^2$
($\dimn = 2$) was created with $M - 1 = 3$ transmitters, $A_1 =0.9$,
$A_2 = 0.8$, $A_3 = 0.7$, $\bm \chi_1=(0.2,0.8)$, $\bm \chi_2=(0.4,0.5)$,
and $\bm \chi_3=(0.8,0.9)$. Using CPQ and enforcing nonnegativity through
$\srn$ virtual measurements per sensor, the {batch semiparametric
  estimate from \eqref{eq28} was computed. The nonparametric part
  adopts a diagonal Gaussian kernel matrix $\bm K ( \bm x_n , \bm
  x_{n'})$ with $k_m (\bm x_n,\bm x_{n'}) = \exp (-\| \bm x_n - \bm
  x_{n'}\|^2/\sigma_m^2)$ on its $m$-th diagonal entry
  (cf. Remark~\ref{remark:kernel}), where $\sigma_\srind^2=0.12$ for
  $\srind=1,2,3,4$. The parametric part is spanned by a basis with
  $N_B=1$ and $\bm B_1({\snind})$ a diagonal matrix whose $(m,m)$-th
  entry is given by $1/(\delta + \| \snind - \bm \chi_m \|^\gamma)$ if
  $m=1,\ldots,M-1$; and 0 if $m=M$.}  The variance $\sigma_\eta^2$ was
set such that about $15\%$ of the measurements contain errors.
Fig.~\ref{fig:maps} shows the true and estimated maps for a particular
realization of sensor locations $\{\snind_n\}$ (represented by
crosses), measurement vectors $\{\msnnmpsnn{\snind_n}{\mpsnind}\}$, and
measurement noise $\eta_p (\bm x_n)$. Each sensor reports $P = 8$
measurements quantized to $b = 5$ bits.  Although every sensor
transmits only $5$ bytes, it is observed that the reconstructed PSD
maps match well with the true ones for all transmitters.

To quantify the estimation performance, the normalized mean-square
error (NMSE), defined as
\begin{align}
\text{NMSE} := \frac{\expected{\|  \chsnn{\snind} - \chestsnn{\snind} \|_2^2}}{\expected{\|  \chsnn{\snind} \|_2^2} }
\end{align}
was employed, where the expectation is taken with respect to the
uniformly distributed $\snind$, measurement vectors
$\{\msnnmpsnn{\snind_n}{\mpsnind}\}$, and measurement noise $\eta_p
(\bm x_n)$. To focus on quantization effects, the region of interest
was the one-dimensional ($d=1$) interval $[0,1]\subset \rfield^1$, and
$\srn-1=4$ transmitters with parameters
$\chi_1=0.1,~\chi_2=0.2,~\chi_3=0.4,~\chi_4=0.8,~A_1=0.8,~A_2=0.9,~A_3=0.8$
and $A_4=0.7$, were considered.

\begin{figure}[t]
\centering
\includegraphics[width=0.46\textwidth]{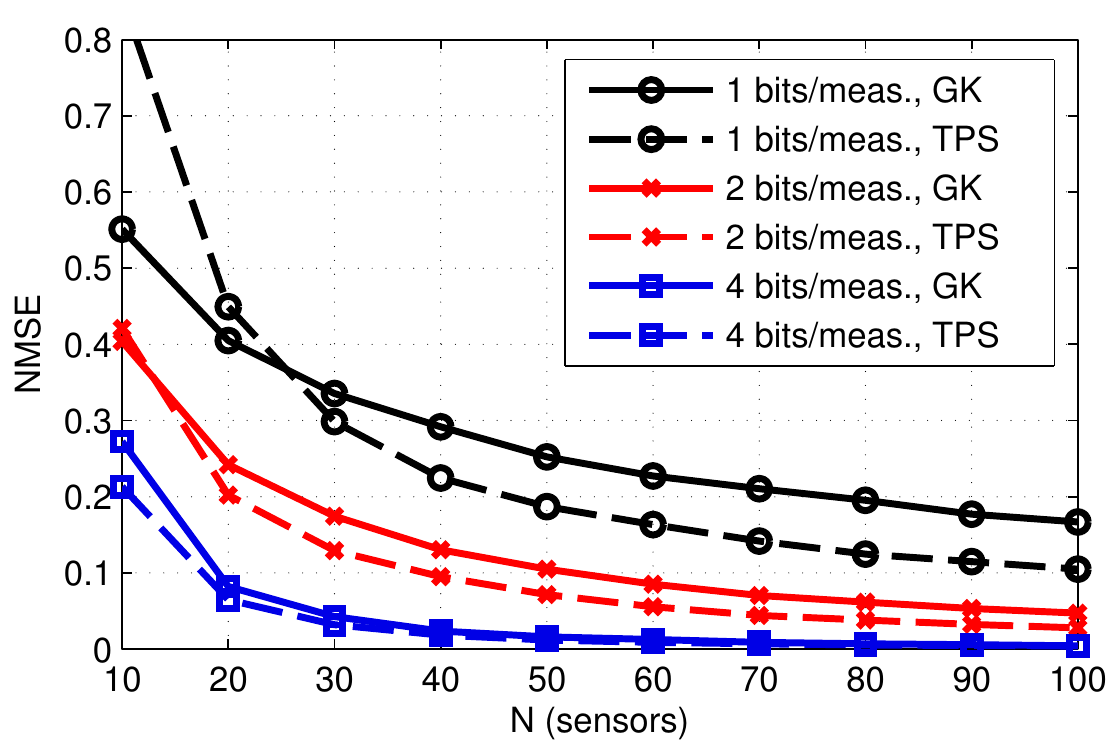}
\caption{NMSE versus $N$ for variable UQ resolution with  
$M = 5$, $P = 5$, $\sigma_\eta^2 =0$, $\lambda =$ $10^{-6}$, and nonnegativity not enforced.}
\label{fig:compression}
\end{figure}

To illustrate the effects of quantization, Fig.~\ref{fig:compression}
depicts the NMSE as a function of $\snn$ for different values of $b$
using both nonparametric and semiparametric estimators. To capture
only quantization effects, no measurement errors were introduced
($\sigma_\eta^2=0$), uniform quantization was used, and $\sigma_s^2$
was set to zero. The nonparametric approach used a diagonal Gaussian
kernel (GK) as in Fig.~\ref{fig:maps}.  It can be seen that the
proposed estimators are consistent in $\snn$. \change{Although, for
  this particular case, TPS mostly outperforms GK-based regression,
  this need not hold in different scenarios since which kernel leads
  to the best performance depends on the propagation environment as
  well as on the field characteristics.}

\begin{figure}[t]
 \centering
 \includegraphics[width=0.46\textwidth]{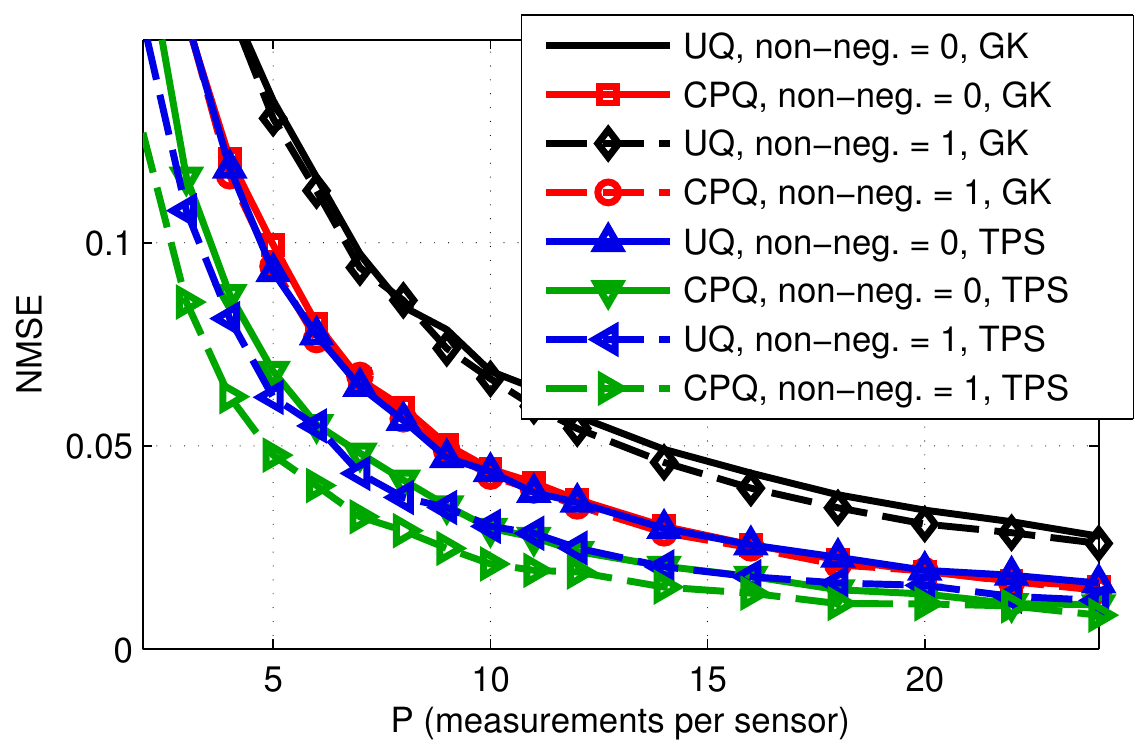}
 \caption{Performance of different quantizers with and without nonnegativity constraints ($M=5$, $N=40$, $b=2$ bits/measurement, $\sigma_\eta^2=0$, and $\lambda=10^{-6}$).}
 \label{fig:extensions}
\end{figure}

\begin{figure}[t]
 \centering
 \includegraphics[width=0.46\textwidth]{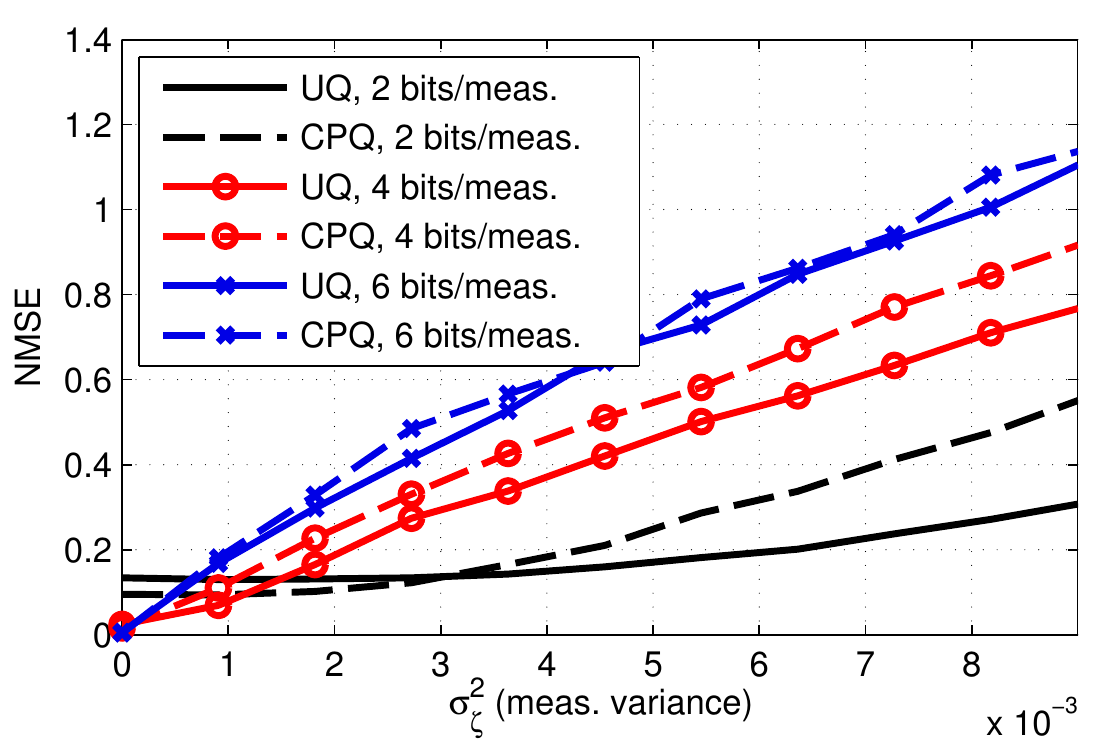}
 \caption{Measurement noise effects ($M=5$, $N=40$, $P=5$, $\lambda=10^{-6}$, GK, and nonnegativity not enforced).}
 \label{fig:measnoise}
\end{figure}

Fig.~\ref{fig:extensions} depicts the NMSE for $N=40$ sensors with a
varying number of measurements $P$ per sensor in different
settings. As expected, the estimates are seen to be inconsistent as
$P$ grows since the number of sensors is fixed and there is no way to
accurately estimate the PSD map far from their vicinity. It is also seen
that TPS regression benefits more from incorporating non-negativity
constraints than the GK-based schemes. Finally, it is observed that
CPQ outperforms UQ. However, \change{as demonstrated by
  Fig.~\ref{fig:measnoise}, UQ leads to a better performance than CPQ
  for this simulation scenario if the noise variance $\sigma_\eta^2$
  is sufficiently large.}  Fig.~\ref{fig:measnoise} further shows that
the effect of measurement noise is more pronounced for larger
$b$. This is intuitive since a given measurement noise variance leads
to more measurement error events. Thus,
$\powflsnnmpsnn{\snind_n}{\mpsnind}$ must be estimated more accurately
under finer quantization.

\begin{figure}[t]
 \centering
 \includegraphics[width=0.49\textwidth]{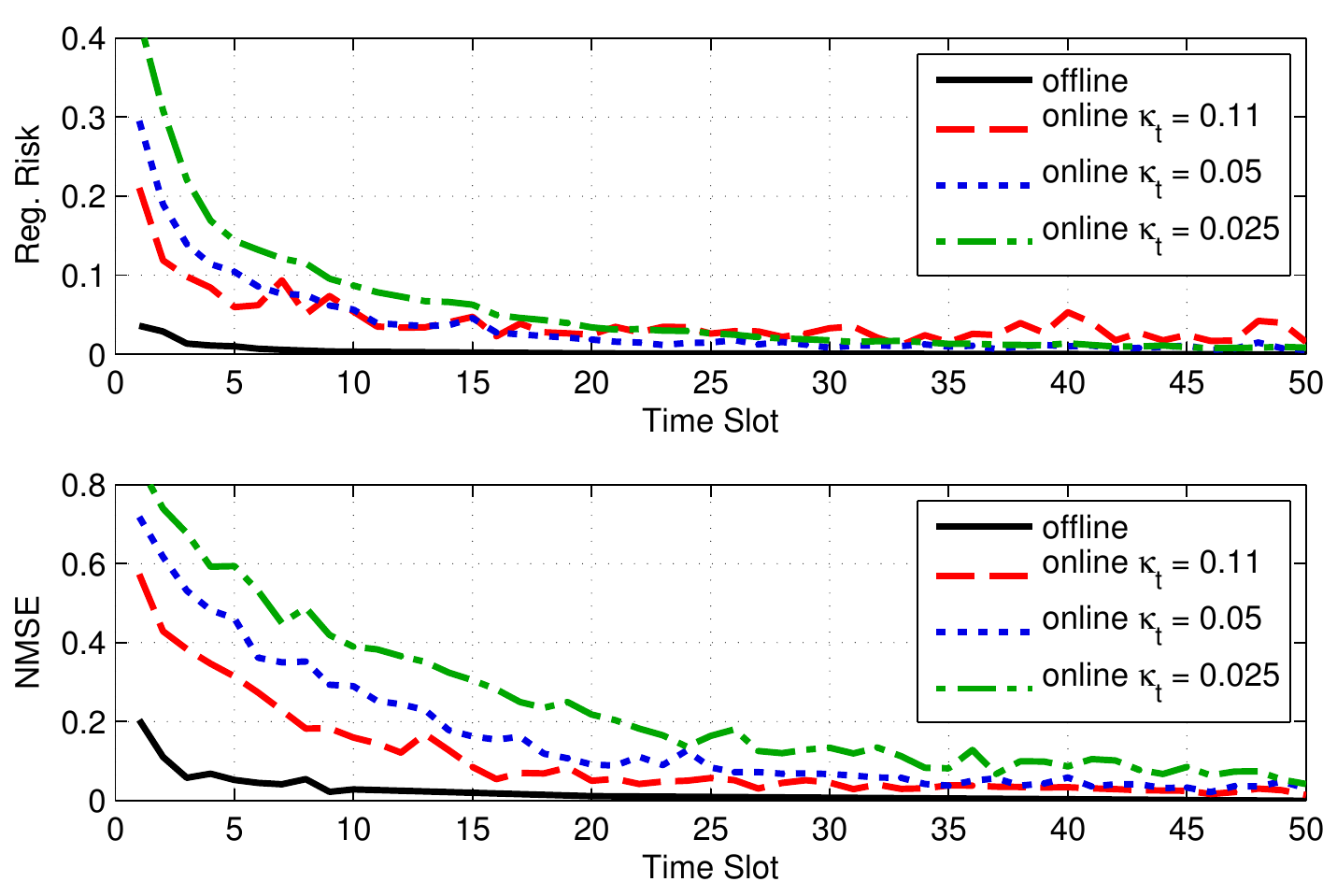}
 \caption{Performance of the online algorithm ($M=5$, $N=30$, $b=4$ bits/measurement, $\sigma_\eta^2=0$, $\lambda=10^{-6}$, GK, and nonnegativity not enforced).}
 \label{fig:online}
\end{figure}

Fig.~\ref{fig:online} depicts the performance of the online algorithm
using the representation in~\eqref{eq:oldec22}. As a benchmark, the
offline (batch) algorithm was run per time slot with all the data
received up to that time slot. The top panel in Fig.~\ref{fig:online}
shows the regularized empirical risks (evaluated per time slot using
the entire set of observations) for different learning rates $\mu_t$.
Common to gradient methods with constant step size, a larger $\mu_t$
speeds up convergence, but also increases the residual error. The
bottom panel depicts the NMSE evolution over time.  Using NMSE as
figure of merit favors greater learning rates over smaller ones.

\section{Conclusions}
\label{sec:con}
This paper introduced a family of methods for nonparametric and
semiparametric estimation of PSD maps using a set of linearly
compressed and possibly quantized power measurements collected by
distributed sensors. To capture different degrees of prior information
without sacrificing flexibility, nonparametric and semiparametric
estimators have been proposed by leveraging the framework of kernel-based
learning. Existing semiparametric regression techniques have been
generalized to vector-valued function estimation and shown to subsume
thin-plate spline regression as a special case. Extensions to multiple
measurements per sensor, non-uniform quantization and non-negativity
constraints have also been introduced. Both batch and online approaches
were developed, thereby offering a performance complexity trade-off.

Future work will be devoted to {kernel
  selection~\cite{gonen2011multiplekernel},} quantizer design, and
alternative types of spectral cartography formats including
construction of channel-gain maps.

\appendices

\section{Derivation of \eqref{eq:dvec}}
\label{app:dvec}

This appendix describes how to obtain $\hat{\tcvec}$ and $\hat{\dvec}$
in \eqref{eq22b} from $(\hbm \alpha,\hbm \beta)$ in
\eqref{eq23}. Because of the change of variable, \eqref{eq23} is not
the dual of \eqref{eq22b}; as a result $\hat{\tcvec}$ and
$\hat{\dvec}$ are not, in general, the Lagrange multipliers of
\eqref{eq23}. As shown next, they can be recovered after relating \eqref{eq23} to the dual of \eqref{eq22b}. The latter is:
\begin{align}
\label{eq:probnpdsp4}
\begin{aligned}
(\check{\tcvec},\check{\alphavec},&\check{\alphasvec})=\argmin_{\tcvec,\alphavec,\alphasvec}\lambda\snn \tcvec^T\proj \Kmat'\proj \tcvec\\
&~-(\obs-\epsilon\bm 1_\snn)^T\alphavec +(\obs+\epsilon \bm 1_\snn)^T\alphasvec\\
\st~~& 2\lambda\snn\proj \Kmat'\proj \tcvec-\proj \Kmat'\cPhimat(\alphavec-\alphasvec)=\bm 0_{\srn\snn}
  \\
&\Psimat^T\cPhimat(\alphavec-\alphasvec)=\bm 0_{\srn\wbasisn} \\
&\bm  \alphavec-\bm 1_\snn\leq \bm 0_\snn ,-\alphavec\leq \bm
  0_\snn,\bm  \alphasvec-\bm 1_\snn\leq \bm 0_\snn ,-\alphasvec\leq \bm 0_\snn.\\
\end{aligned}
\end{align}
The presence of $\tcvec$ in both \eqref{eq22b} and
\eqref{eq:probnpdsp4} is owing to the fact that $\proj \Kmat'\proj$ is not
invertible.  Taking into account that \eqref{eq22b} is indeed the dual
of \eqref{eq:probnpdsp4}, it can be shown that $\hat{\dvec}$ is the
Lagrange multiplier $\cbm \mu_2$ associated with the second equality
constraint of \eqref{eq:probnpdsp4}, whereas $\hat{\tcvec}
=\check{\tcvec}$. Thus, it remains only to express $\cbm \mu_2$ and
$\check{\tcvec}$ in terms of the solution to \eqref{eq23}.

From the second equality constraint in \eqref{eq:probnpdsp4}, it holds
for $(\alphavec,\alphasvec)$ feasible that
$\cPhimat(\alphavec-\alphasvec)=\proj\cPhimat(\alphavec-\alphasvec)$. Then,
the first equality constraint in \eqref{eq:probnpdsp4} can be replaced
with
\begin{align}
\label{eq:puff}
2\lambda\snn\proj \Kmat'\proj \tcvec
-\proj \Kmat'\proj\cPhimat(\alphavec-\alphasvec)=\bm 0_{\srn\snn}.
\end{align}
Solving \eqref{eq:puff} for $\tcvec$ produces
\begin{align}
\label{eq:tcvecopt}
\tcvec = \frac{1}{2\lambda\snn} \cPhimat(\alphavec-\alphasvec) +
\bm \nu(
\proj \Kmat'\proj)
\end{align}
where $\bm \nu(\bm A)$ denotes any vector in the null space of $\bm
A$.  Substituting \eqref{eq:tcvecopt} into \eqref{eq:probnpdsp4},
one recovers \eqref{eq23}.

Clearly,  problems \eqref{eq:probnpdsp4} and \eqref{eq23} are
equivalent in the sense that if $(\hat\alphavec,\hat\alphasvec)$
solves the latter, then 
\begin{align}
\label{eq:equps}
&\check\tcvec = \frac{1}{2\lambda\snn}
  \cPhimat(\hat\alphavec-\hat\alphasvec),
~~\check\alphavec = \hat \alphavec,~~\check\alphasvec = \hat \alphasvec
\end{align}
solve the former. However, their Lagrange multipliers differ due to the
transformation introduced. A possible means of establishing their
relation is to compare the KKT conditions of both problems. In
particular, let $\hbm \mu_2$, $\hbm \nu_1$, $\hbm \nu_2$, $\hbm \nu_3$,
$\hbm \nu_4$ denote the Lagrange multipliers corresponding to
$(\hat\alphavec,\hat\alphasvec)$, associated with the constraints of
\eqref{eq23} in the same order listed here; then the multipliers
\begin{align}
\begin{aligned}
&\cbm \mu_1=-\check\tcvec,~~\cbm \mu_2=\hbm \mu_2- (\Psimat^T\Psimat)\inv
\Psimat^T \Kmat'  \check\tcvec,\label{eq:middle}\\
&\cbm  \nu_1=\hbm \nu_1,~~\cbm \nu_2=\hbm \nu_2,~~\cbm \nu_3=\hbm \nu_3,~~\cbm \nu_4=\hbm \nu_4
\end{aligned}
\end{align}
correspond to the solution of \eqref{eq:probnpdsp4} given by
\eqref{eq:equps}. Recalling that $\hat \dvec = \cbm \mu_2$ one readily
arrives at \eqref{eq:dvec}.

\section{Efficient Implementation of A Special Case}
In practice, one may effectively ignore the dependencies between the
entries $\chsrnsnn{\srind}{\snind},~\srind=1,\ldots,\srn$, by using
a diagonal kernel $\Kmat(\snind,\snindgen)$ and diagonal basis
functions $\wbasiswbasisnsnn{\wbasisind}{\snind}$. Furthermore, one
may be interested in modeling all entries identically,  as
in TPS regression. Then, both the kernel and basis
functions become scaled identity matrices; that is, for
certain scalar functions $K(\bm x, \bm x')$ and $B_\nu(\bm x)$, one has
\begin{align}
\Kmat(\snind,\bm x') &:= K(\snind,\bm x')\bm I_\srn\\ \wbasiswbasisnsnn{\wbasisind}{\snind}
&:=\wbasiswbasisnsnns{\wbasisind}{\snind}\bm
I_\srn,~\wbasisind=1,\ldots,\wbasisn.
\end{align}
Thus, upon defining $\Kmatsm\in \rfield^{\snn\times\snn}$ with $(n,n')$-entry equal to $K(\snind_n,\snind_{n'})$, and $\Psimatsm\in\rfield^{\snn\times\wbasisn}$ with $(n,\nu)$-entry equal to $\wbasiswbasisnsnns{\nu}{\snind_n}$, matrices $\bm K$ and $\bm B$ can be written as
\begin{align}
\label{eq:kmatkron}
\Kmat = \Kmatsm\otimes \bm I_\srn\\
\label{eq:psimatkron}
\Psimat = \Psimatsm\otimes \bm I_\srn.
\end{align}
Then, the computation of certain matrices involved in the proposed
algorithms can be done efficiently as described next.

Start by constructing a selection matrix $\Smat$ containing ones
and zeros such that $\bm A \odot \bm B = (\bm A \otimes \bm B)\Smat$,
and define $\bar{\bm \Phi}_0 := (\bm I_N \otimes {\bm
  1}_P^T) \odot \bar{\bm \Phi}$
(c.f. Sec.~\ref{sec:emms}) and~\eqref{eq:kmatkron} to obtain
\begin{align}
\begin{aligned}
\cbPhimat^T\Kmat &=
[(\bm I_\snn\otimes \bm
  1_\mpsnn^T)\odot \bPhimat]^T(\Kmatsm\otimes \bm I_\srn)\\
 &=
\Smat^T[(\bm I_\snn\otimes \bm
  1_\mpsnn)\otimes \bPhimat^T](\Kmatsm\otimes \bm I_\srn)\\
 &=
\Smat^T[(\bm I_\snn\otimes \bm
  1_\mpsnn)\Kmatsm\otimes \bPhimat^T]\\
 &=
\Smat^T[(\Kmatsm\otimes \bm
  1_\mpsnn)\otimes \bPhimat^T]\\
 &=
[(\Kmatsm\otimes \bm
  1_\mpsnn^T)\odot \bPhimat]^T.
\end{aligned}
\end{align}
Likewise,  one can verify that
\begin{align}
\cbPhimat^T\Psimat &= [(\Psimatsm^T\otimes \bm 1_\mpsnn^T)\odot
  \bPhimat]^T.
\end{align}
Using the Kronecker product properties, $\proj$ can be written as
\begin{align}
\begin{aligned}
\proj& = \bm I_{\srn\snn} - (\Psimatsm(\Psimatsm^T\Psimatsm)\inv
\Psimatsm^T) \otimes \bm I_\srn\\
& = (\bm I_{\snn} - \Psimatsm(\Psimatsm^T\Psimatsm)\inv
\Psimatsm^T) \otimes \bm I_\srn = \projz \otimes \bm I_\srn
\end{aligned}
\end{align}
where $\projz:=\bm I_{\snn} - \Psimatsm(\Psimatsm^T\Psimatsm)\inv \Psimatsm^T$.  
Also, from~\eqref{eq:tbbRdef}
\begin{align}
\label{eq:trmatc}
\begin{aligned}
\tbRmat&= \cbPhimat^T(\projz \otimes \bm
I_\srn)(\Kmatsm\otimes \bm I_\srn)(\projz \otimes \bm I_\srn)
\cbPhimat\\
&= \cbPhimat^T(\projz \Kmatsm \projz \otimes \bm I_\srn)
\cbPhimat\\
&= ((\bm I_\snn\otimes \bm 1_\mpsnn^T)\odot \bPhimat)^T(\projz \Kmatsm \projz \otimes \bm I_\srn) \cdot ((\bm I_\snn\otimes \bm 1_\mpsnn^T)\odot \bPhimat)\\
&= \Smat^T((\bm I_\snn\otimes \bm 1_\mpsnn^T)\otimes \bPhimat)^T(\projz \Kmatsm \projz \otimes \bm I_\srn)
\\&~~~~\cdot ((\bm I_\snn\otimes \bm 1_\mpsnn^T)\otimes \bPhimat)\Smat\\
&= \Smat^T[(\bm I_\snn\otimes \bm 1_\mpsnn)\projz \Kmatsm \projz(\bm
I_\snn\otimes \bm 1_\mpsnn^T) \otimes \bPhimat^T  \bPhimat]\Smat\\
&= (\bm I_\snn\otimes \bm 1_\mpsnn)\projz \Kmatsm \projz(\bm
I_\snn\otimes \bm 1_\mpsnn^T) \circ \bPhimat^T  \bPhimat\\
&= (\projz \Kmatsm \projz \otimes \bm 1_\mpsnn \bm 1_\mpsnn^T) \circ \bPhimat^T  \bPhimat.
\end{aligned}
\end{align}
Finally, $\dvec$ can be obtained as (cf. Sec.~\ref{sec:emms}):
\begin{align}
\dvec = \bar{\bm \mu} - [ (\Psimatsm^T\Psimatsm)\inv \Psimatsm^T \Kmatsm \otimes \bm I_\srn]  \hat{\bar{\cvec}}.
\end{align}

\section{Proof of Theorem~\ref{thm:conv}}
\label{sec:proof:thm:conv}
{The following proof extends that
  of~\cite[Thm. 1]{audiffren2013operatorvalued} which cannot
  accommodate the instantaneous cost from \eqref{eq:instregerrdef}
 when $\mathcal{L}=\mathcal{L}_{1\epsilon}$  since
   $\mathcal{L}_{1\epsilon}$  is not differentiable and $\bm \phi(\bm
       x_\nu)$ is not  constant over $\nu$. 
} 
{
Expanding the norms into inner products and employing 
\eqref{eq:upd}, one finds that
\begin{align}
\nonumber&||\bm l^{(t)}-\bm g||_\mathcal{H}^2
-
||\bm l^{(t+1)}-\bm g||_\mathcal{H}^2\\
\nonumber&=
-||\bm l^{(t+1)}-\bm l^{(t)}||_\mathcal{H}^2
-2\langle \bm l^{(t+1)}-\bm l^{(t)},
\bm l^{(t)}-\bm g
\rangle_\mathcal{H}^2\\
&=
-\mu_t^2||\partial_{\bm l}
\instregerr(\w\atstn{\stind},\msnn{\snindtt},\snindtt,\obssnn{\snindtt})
||_\mathcal{H}^2\label{eq:twotermstobound}\\
\nonumber&+2\langle \mu_t\partial_{\bm l}\instregerr(\w\atstn{\stind},\msnn{\snindtt},\snindtt,\obssnn{\snindtt}),
\bm l^{(t)}-\bm g
\rangle_\mathcal{H}.
\end{align}
To bound the norm in \eqref{eq:twotermstobound}, apply the triangle inequality to
\eqref{eq:gradrinst} to obtain: 
\begin{align*}
&||\partial_{\w} \instregerr(\w,\m(\snind_\nu),\snind_\nu,\obsel(\snind_\nu)) ||_{\mathcal{H}}
\\ &\leq |\mathcal{L}'_{1\epsilon}(\obsel(\snind_\nu)-\m^T(\snind_\nu)
\w(\snind_\nu))|~
||\bm K(\cdot, \snind_\nu)\m(\snind_\nu)||_{\mathcal{H}}
+2\lambda||\w||_{\mathcal{H}}.
\nonumber
\end{align*}
Since $\mathcal{L}_{1\epsilon}$ is 1-Lipschitz,
$\mathcal{L}'_{1\epsilon}(e)\leq 1~\forall e$. Moreover,  the
reproducing property (cf. Sec.~\ref{sec:nprnq}) implies that $||\bm
K(\cdot,\snind_\nu)\m(\snind_\nu)||_{\mathcal{H}}
= ({\m^T(\snind_\nu) \bm K( \snind_\nu, \snind_\nu)\m(\snind_\nu)})^{1/2}
< \bar{\lambda}||\m(\snind_\nu)||$, and hence
\begin{align}
\label{eq:boundfirstterm1}
&||\partial_{\w} \instregerr(\w,\m(\snind_\nu),\snind_\nu,\obsel(\snind_\nu)) ||_{\mathcal{H}}
\leq \bar{\lambda}||\m(\snind_\nu)||
+2\lambda||\w||_{\mathcal{H}}.
\end{align}
Similarly, from \eqref{eq:wstnstind} and the triangular inequality, one
finds that
\begin{align*}
||\w\atstn{\stind+1}&||_{\mathcal{H}}\leq
(1-2\step\lambda)||\w\atstn{\stind}||_{\mathcal{H}}+\step%
\bar{\lambda}||\m(\snind_\nu)||.
\end{align*}
Recalling that $\w\atstn{1}=\bm 0$, it is simple to show by induction
that $||\w\atstn{\stind}||_{\mathcal{H}}\leq U\define {\bar\lambda
  \bar{\varphi}}/{2\lambda}$ for all $t$. This fact and $
||\m({\snindtt})||_2\leq \mbound$  applied to 
\eqref{eq:boundfirstterm1} produce
\begin{align}
\label{eq:boundfirstterm}
&||\partial_{\w} \instregerr(\w,\m(\snind_\nu),\snind_\nu,\obsel(\snind_\nu)) ||_{\mathcal{H}}
\leq \bar{\lambda}  \bar{\varphi}
+2\lambda U = 2\bar{\lambda}  \bar{\varphi}.
\end{align}
for all $\bm l$.  On the other hand, the last term in
\eqref{eq:twotermstobound} can be bounded by invoking the definition of
subgradient:
\begin{align}
\label{eq:boundsecondterm}
&\langle \partial_{\bm l}\instregerr(\w\atstn{\stind},\msnn{\snindtt},\snindtt,\obssnn{\snindtt}),
\bm l^{(t)}-\bm g
\rangle_\mathcal{H}
\\&\geq
\instregerr(\w\atstn{\stind},\msnn{\snindtt},\snindtt,\obssnn{\snindtt})
-
\instregerr(\bm g,\msnn{\snindtt},\snindtt,\obssnn{\snindtt}).\nonumber
\end{align}
Combining \eqref{eq:boundfirstterm} and \eqref{eq:boundsecondterm}
with \eqref{eq:twotermstobound} results in 
\begin{align*}
&||\bm l^{(t)}-\bm g||_\mathcal{H}^2
-
||\bm l^{(t+1)}-\bm g||_\mathcal{H}^2\\
&\geq
-4\mu_t^2\bar{\lambda}^2  \bar{\varphi}^2
-2\mu_t\big[
\instregerr(\bm g,\msnn{\snindtt},\snindtt,\obssnn{\snindtt})\nonumber
\\&
- \instregerr(\w\atstn{\stind},\msnn{\snindtt},\snindtt,\obssnn{\snindtt}\big].
\end{align*}
Adapting the proof
of~\cite[Prop. 3.1(iii)]{audiffren2013operatorvalued}, it can be shown
that if $\bm g\in \mathcal{H}$ equals the value of $\bm l$ attaining
the infimum on the right hand side of \eqref{eq:regretbound}, then
$||\bm g||_\mathcal{H}\leq U$. For such a $\bm g$ one has
\begin{align}
&\frac{1}{\mu_t}||\bm l^{(t)}-\bm g||_\mathcal{H}^2
-
\frac{1}{\mu_{t+1}}||\bm l^{(t+1)}-\bm g||_\mathcal{H}^2\\
&=\frac{1}{\mu_t}\left[||\bm l^{(t)}-\bm g||_\mathcal{H}^2
-||\bm l^{(t+1)}-\bm g||_\mathcal{H}^2\right]
\\&-\left(\frac{1}{\mu_{t+1}}-\frac{1}{\mu_t}\right)||\bm l^{(t+1)}-\bm g||_\mathcal{H}^2\\ %
&\geq
-4\mu_t\bar{\lambda}^2  \bar{\varphi}^2
-2\big[
\instregerr(\bm g,\msnn{\snindtt},\snindtt,\obssnn{\snindtt})\nonumber
\\&
- \instregerr(\w\atstn{\stind},\msnn{\snindtt},\snindtt,\obssnn{\snindtt}\big]
+\left(\frac{1}{\mu_t}-\frac{1}{\mu_{t+1}}\right)4U^2
\end{align}
since $||\bm l^{(t+1)}-\bm g||_\mathcal{H}\leq 2U$. Summing for
$t=1,\ldots,T$ and applying $\sum_{t=1}^T\mu_t \leq 2\mu
\sqrt{T}$~\cite{kivinen2004online} together with $\sqrt{T+1}-1\leq
\sqrt{T}$ the result in~\eqref{eq:regretbound} follows readily.
}

\bibliographystyle{IEEEtranS}
\bibliography{my_bibliography,local_bibliography}

%

\begin{IEEEbiography}[{\includegraphics[width=1in,height=1.25in,clip,keepaspectratio]{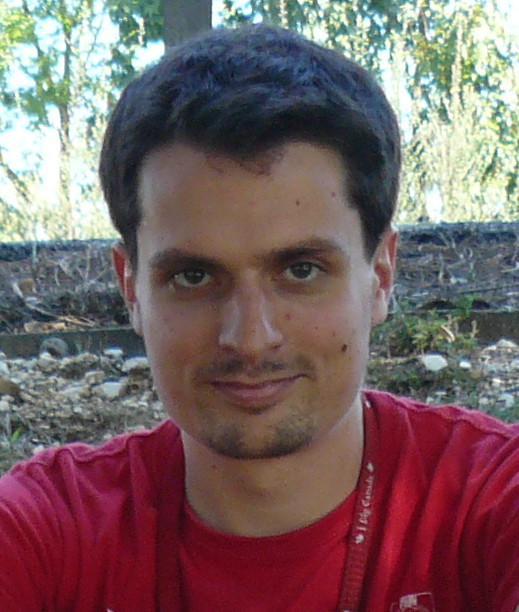}}]{D. Romero
  (M'16)} received his M.Sc. and Ph.D. degrees in Signal Theory and
  Communications from the University of Vigo, Spain, in 2011 and 2015,
  respectively. From Jul. 2015 to Nov. 2016, he was a post-doctoral
  researcher with the Digital Technology Center and Department of
  Electrical and Computer Engineering, University of Minnesota,
  USA. In Dec. 2017, he joined the Department of Information and
  Communication Technology, University of Agder, Norway, as an
  associate professor. His main research interests lie in the areas of
  signal processing, communications, and machine learning.
\end{IEEEbiography}

\vfill

\begin{IEEEbiography}[{\includegraphics[width=1in,height=1.25in,clip,keepaspectratio]{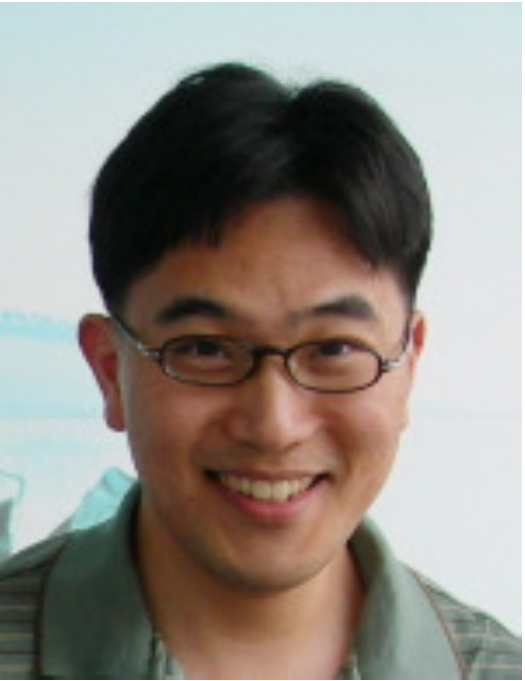}}]{Seung-Jun Kim (SM'12)} received his B.S. and M.S. degrees from Seoul National University in Seoul, Korea in 1996 and 1998, respectively, and his Ph.D. from the University of California at Santa Barbara in 2005, all in electrical engineering. From 2005 to 2008, he worked for NEC Laboratories America in Princeton, New Jersey, as a Research Staff Member. He was with Digital Technology Center and Department of Electrical and Computer Engineering at the University of Minnesota during 2008-2014, where his final title was Research Associate Professor. In August 2014, he joined Department of Computer Science and Electrical Engineering at the University of Maryland, Baltimore County, as an Assistant Professor. His research interests include statistical signal processing, optimization, and machine learning, with applications to wireless communication and networking, future power systems, and (big) data analytics. He is serving as an Associate Editor for IEEE Signal Processing Letters. 
\end{IEEEbiography}

\begin{biography}[{\includegraphics[width=1in,height=1.25in,clip,keepaspectratio]{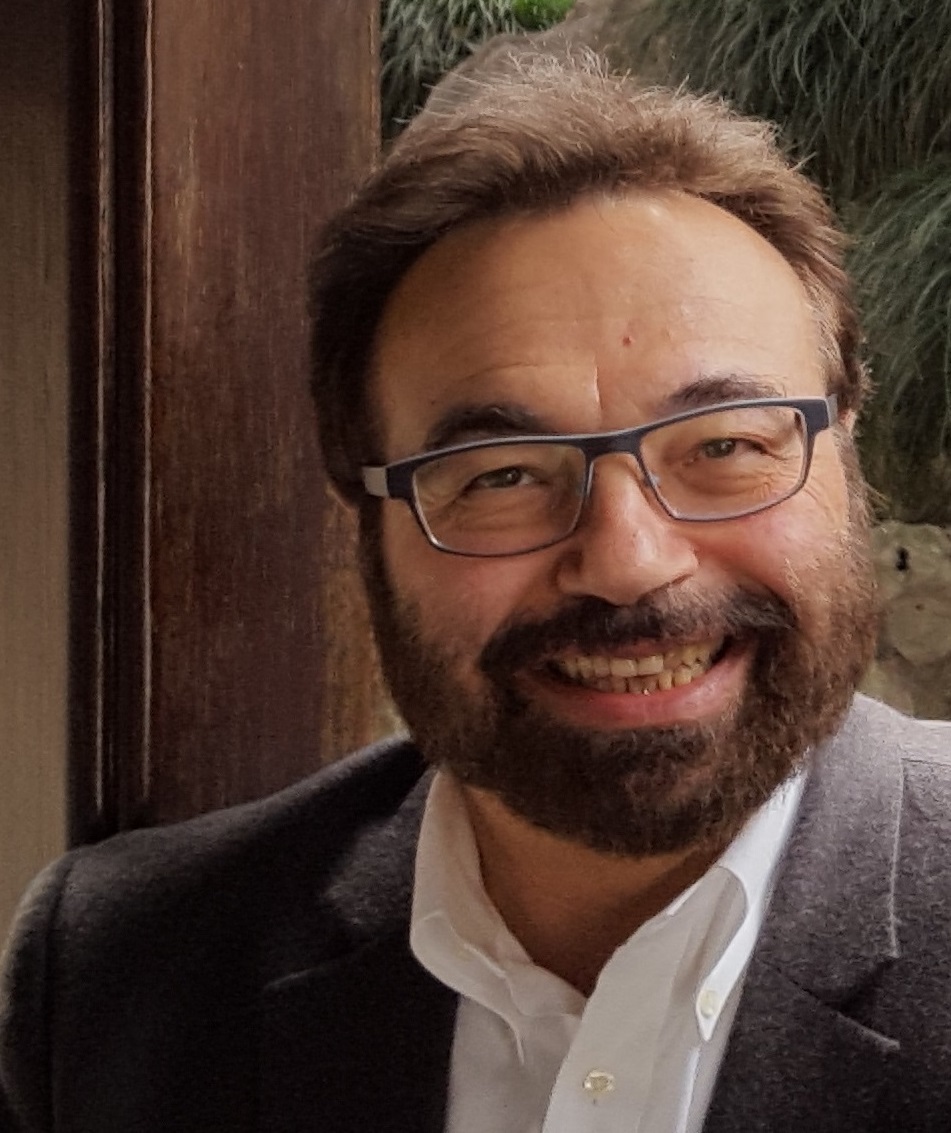}}]{G. B.  Giannakis (Fellow'97)} received his
  Diploma in Electrical Engr. from the Ntl. Tech. Univ. of Athens,
  Greece, 1981. From 1982 to 1986 he was with the Univ. of Southern
  California (USC), where he received his MSc. in Electrical
  Engineering, 1983, MSc.  in Mathematics, 1986, and Ph.D. in
  Electrical Engr., 1986. He was with the University of Virginia from
  1987 to 1998, and since 1999 he has been a professor with the
  Univ. of Minnesota, where he holds an Endowed Chair in Wireless
  Telecommunications, a University of Minnesota McKnight Presidential
  Chair in ECE, and serves as director of the Digital Technology
  Center.

  His general interests span the areas of communications, networking
  and statistical signal processing - subjects on which he has
  published more than 400 journal papers, 680 conference papers, 25
  book chapters, two edited books and two research monographs (h-index
  123). Current research focuses on learning from Big Data, wireless
  cognitive radios, and network science with applications to social,
  brain, and power networks with renewables. He is the (co-) inventor
  of 28 patents issued, and the (co-) recipient of 8 best paper awards
  from the IEEE Signal Processing (SP) and Communications Societies,
  including the G. Marconi Prize Paper Award in Wireless
  Communications. He also received Technical Achievement Awards from
  the SP Society (2000), from EURASIP (2005), a Young Faculty Teaching
  Award, the G. W. Taylor Award for Distinguished Research from the
  University of Minnesota, and the IEEE Fourier Technical Field Award
  (2015). He is a Fellow of EURASIP, and has served the IEEE in a
  number of posts, including that of a Distinguished Lecturer for the
  IEEE-SP Society.
\end{biography}

\begin{IEEEbiography}[{\includegraphics[width=1in,height=1.25in,clip,keepaspectratio]{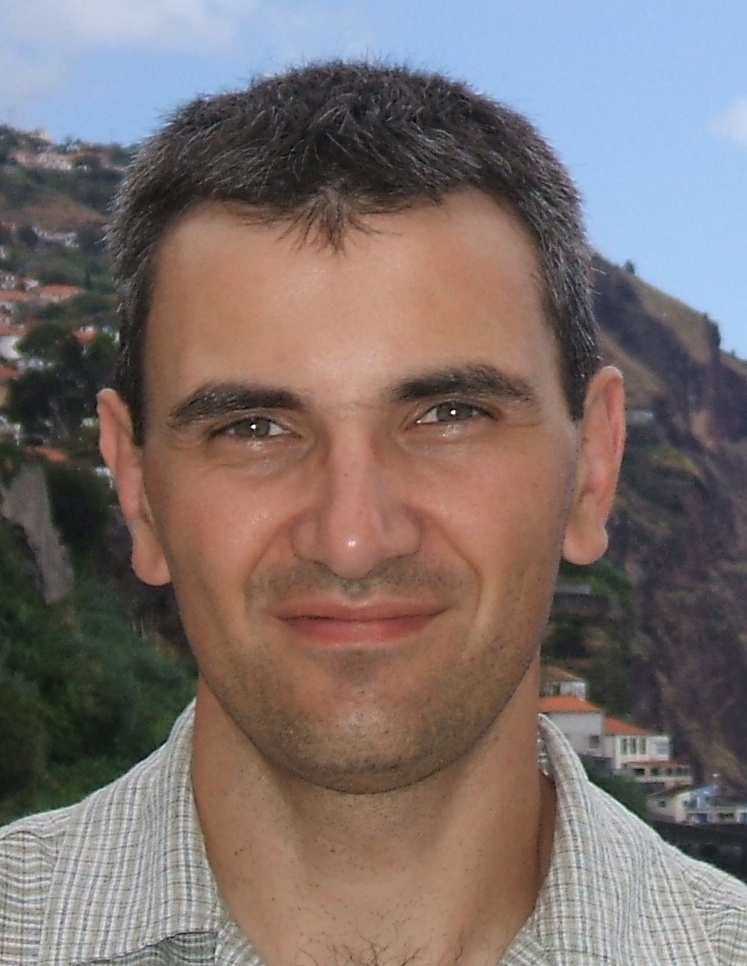}}]{Roberto L\'opez-Valcarce 
(S'95-M'01)} received the Telecommunication Engineering degree from
  the University of Vigo, Vigo, Spain, in 1995, and the M.S. and
  Ph.D. degrees in electrical engineering from the University of Iowa,
  Iowa City, in 1998 and 2000, respectively. During 1995 he was a
  Project Engineer with Intelsis. He was a {\em Ram\'on y Cajal}
  Postdoctoral Fellow of the Spanish Ministry of Science and
  Technology from 2001 to 2006. During that period, he was with the
  Signal Theory and Communications Department, University of Vigo,
  where he currently is an Associate Professor. His main research
  interests include adaptive signal processing, digital
  communications, and sensor networks, having coauthored over 50
  papers in leading international journals. He holds several patents
  in collaboration with industry.

Dr. L\'opez-Valcarce was a recipient of a 2005 Best Paper Award of the IEEE Signal Processing Society. He served as an Associate Editor of the IEEE TRANSACTIONS ON SIGNAL PROCESSING from 2008 to 2011, and as a member of the IEEE Signal Processing for Communications and Networking Technical Committee from 2011 to 2013.
\end{IEEEbiography}

\vfill




\end{document}